\documentclass[aps,physrev,reprint,superscriptaddress]{revtex4-2}

\usepackage{mathtools}
\usepackage{amssymb}
\usepackage[mathscr]{euscript}
\usepackage{siunitx}
\usepackage{accents}
\usepackage{bm}
\usepackage[utf8]{inputenc}
\usepackage[T1]{fontenc}
\usepackage{xcolor}
\usepackage{graphicx}
\usepackage{algorithm}
\usepackage[noend]{algpseudocode}
\usepackage[hidelinks]{hyperref}

\begin{document}

\newfloat{algorithm}{htbp}{loa}

\title{Quasicrystal Architected Nanomechanical Resonators via Data-Driven Design}

\author{Kawen Li}
\affiliation{Department of Precision and Microsystems Engineering, Delft University of Technology, Delft, 2628CD, The Netherlands}

\author{Hangjin Cho}
\affiliation{Department of Mechanical Engineering, Pohang University of Science and Technology, Pohang, 37673, Republic of Korea}

\author{Richard Norte}
\email{r.a.norte@tudelft.nl}
\affiliation{Department of Precision and Microsystems Engineering, Delft University of Technology, Delft, 2628CD, The Netherlands}

\author{Dongil Shin}
\email{dongilshin@postech.ac.kr}
\affiliation{Department of Mechanical Engineering, Pohang University of Science and Technology, Pohang, 37673, Republic of Korea}

\date{\today}

\begin{abstract}
From butterfly wings to remnants of nuclear detonation, aperiodic order repeatedly emerges in nature, often exhibiting reduced sensitivity to boundaries and symmetry constraints. 
Inspired by this principle, a paradigm shift is introduced in nanomechanical resonator design from periodic to aperiodic structures, focusing on a special class: quasicrystals (QCs). 
Although soft clamping enabled by phononic stopbands has become a central strategy for achieving high-$Q_m$ nanomechanical resonators, its practical realization has been largely confined to periodic phononic crystals, where band structure engineering is well established. 
The potential of aperiodic architectures, however, has remained largely unexplored, owing to their intrinsic complexity and the lack of systematic approaches to identifying and exploiting stopband behavior.
Here we demonstrate that soft clamping can be realized in quasicrystal architectures and that high-$Q_m$ nanomechanical resonators can be systematically predicted through a data-driven design framework.
As a representative demonstration, the 12-fold QC-based resonator obtained numerically a quality factor $Q_m \sim 10^7$ and an effective mass of sub-nanograms at MHz frequencies, corresponding to an exceptional force sensitivity of $26.4$~aN/$\sqrt{\text{Hz}}$ compared to previous 2D phononic crystals. 
These results establish QCs as a robust platform for next-generation nanomechanical resonators and open a new design regime beyond periodic order. 
\end{abstract}


\maketitle

\section{\label{sec:1} Introduction}
Recent advances in nanofabrication have enabled silicon-based nanomechanical resonators with ultra-low dissipation~\cite{engelsen_ultrahigh-quality-factor_2024}, corresponding to exceptionally high mechanical quality factors ($Q_m$). 
Such resonators are essential for quantum-limited sensing~\cite{halg_membrane-based_2021,kippenberg_cavity_2008,mason_continuous_2019,krause_high-resolution_2012} and macroscopic quantum experiments~\cite{yang_mechanical_2024,satzinger_quantum_2018,oconnell_quantum_2010}, owing to their ability to isolate mechanical motion from environmental noise. 
While advances in material quality~\cite{xu_high-strength_2024} and fabrication techniques enabled extreme aspect ratios~\cite{cupertino_centimeter-scale_2024} and high tensile prestress played a central role, the past decade has also seen increasing emphasis on exploiting fundamental structural design strategies to further suppress dissipation~\cite{shin_spiderweb_2022,fedorov_fractal-like_2020,shi_topology_2026,algra_dissipation_2025,hoj_ultra-coherent_2021,hoj_ultracoherent_2024}. 
A key challenge in such highly stressed, slender structures is that dissipation is often dominated by bending losses concentrated in regions of large curvature~\cite{schmid_damping_2011}, motivating the development of design principles that mitigate this loss channel. 
In this context, periodic phononic crystal (PnC) architectures have emerged as a dominant approach~\cite{tsaturyan_demonstration_2014,tsaturyan_ultracoherent_2017,ghadimi_radiation_2017,ghadimi_elastic_2018,reetz_analysis_2019,cupertino_centimeter-scale_2024}, where mechanical stopbands formed via Bragg scattering~\cite{kushwaha_theory_1994} enable the realization of soft-clamped resonance modes that spatially confine vibrational energy away from lossy supports. 
Despite its success in achieving record-high $Q_m$, stopband-enabled soft clamping remains largely restricted to periodic, unit-cell-based architectures.

\begin{figure}[t]
\includegraphics[width=\linewidth]{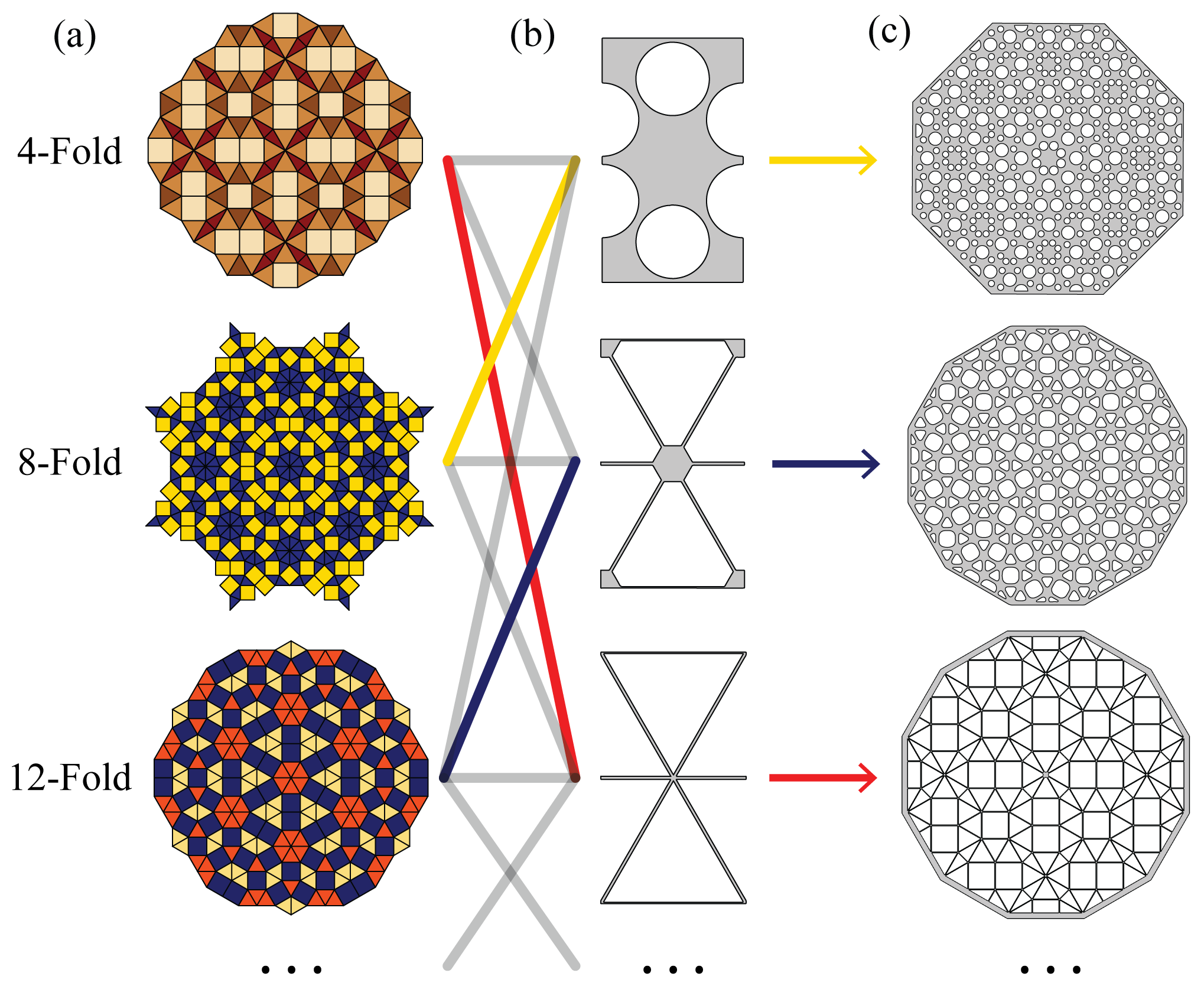}
\caption{\label{fig:introduction} 
\textbf{Overview of the quasicrystal-enabled resonator design space}
(a) Quasicrystal tilings with 4-, 8-, and 12-fold symmetry.
(b) Structural motifs inspired by phononic crystal unit cell designs.
(c) Resulting quasicrystal nanomechanical resonator geometries enabled by aperiodic order.
}
\end{figure}

This fundamental reliance on translational symmetry raises a critical question: is strict periodicity truly required for stopband formation with soft clamping, or can alternative forms of structural order provide comparable or even enhanced wave-isolation capabilities? 
Beyond strictly periodic architectures, quasicrystals (QCs) represent a class of aperiodic structures that lack translational periodicity yet retain long-range order through noncrystallographic rotational symmetries, allowing higher-order symmetries such as eight-, ten-, and twelvefold symmetries~\cite{levine_quasicrystals_1984,levine_quasicrystals_1986,zoorob_complete_2000}. 
The formation of such aperiodic order has been explored in both theoretical~\cite{fayen_quasicrystal_2024,varela-rosales_computational_2025} and experimental~\cite{bindi_electrical_2023,bindi_accidental_2021} contexts, primarily in terms of geometric realization and structural properties.
When combined with established structural design strategies, QC architectures open a largely unexplored design space for extending stopband-enabled soft clamping beyond periodic PnC, offering a fundamentally different route to wave isolation that does not rely on translational symmetry.
Fig.~\ref{fig:introduction} illustrates representative QC resonator geometries enabled by this aperiodic order.

Freed from the constraints of periodicity, individual components in QC-based architectures can, in principle, be tuned to achieve tailored mechanical responses. 
However, this flexibility weakens the intuitive link between geometry, stopband characteristics, and resonator performance, making manual design and tuning impractical.
In contrast to periodic systems, where stopbands can be systematically determined using unit-cell analysis and Bloch’s theorem~\cite{kushwaha_theory_1994}, such approaches are fundamentally inapplicable to quasi-periodic architectures. 
As a result, prior studies have relied on periodic approximants to estimate QC stopbands in both photonic~\cite{rechtsman_optimized_2008,florescu_complete_2009,della_villa_band_2005,chan_photonic_1998} and mechanical~\cite{xia_topological_2020,rosa_edge_2019,pal_topological_2019} fields.
While these approaches provide valuable insight into global stopband characteristics, additional considerations are required for soft-clamped nanomechanical resonator design, which relies on localized defect modes within clean stopbands.
Recent efforts exploiting fold-symmetry conditions have improved defect-mode detection in elastic and acoustic systems; however, they still require manual identification of stopband boundaries, limiting scalability and introducing subjectivity~\cite{beli_mechanics_2021}.
These challenges underscore the need for an automated and systematic framework for the design and optimization of QC-based nanomechanical resonators, which to the best of our knowledge does not yet exist.

In this work, we introduce a data-driven design approach that enables stopband-based soft clamping in QC-based nanomechanical resonators.
Our approach directly identifies stopbands in quasi-periodic architectures and systematically exploits them to realize high-$Q_m$ soft-clamped defect modes, without relying on translational periodicity or unit-cell band-structure concepts.
Fig.~\ref{fig:introduction} shows representative QC-based nanomechanical resonator geometries designed in this work, spanning different rotational symmetries and structural motifs.
Beyond extending soft clamping to aperiodic architectures, this work reveals a mechanical design regime that combines efficient dissipation dilution with large effective mass compared to 1D architectures, thereby bridging characteristics that were traditionally separated in periodic nanomechanical systems.
Together, this work establishes a practical pathway for translating the structural complexity of quasicrystals into stopband-enabled, soft-clamped nanomechanical resonators with unique mechanical performance.

\section{\label{sec:2} Results}
\subsection{\label{sec:2.A} Stopband Formation and Scaling in Quasicrystal Nanomechanical Resonators}

\begin{figure}[b]
    \centering
    \includegraphics[width=\linewidth]{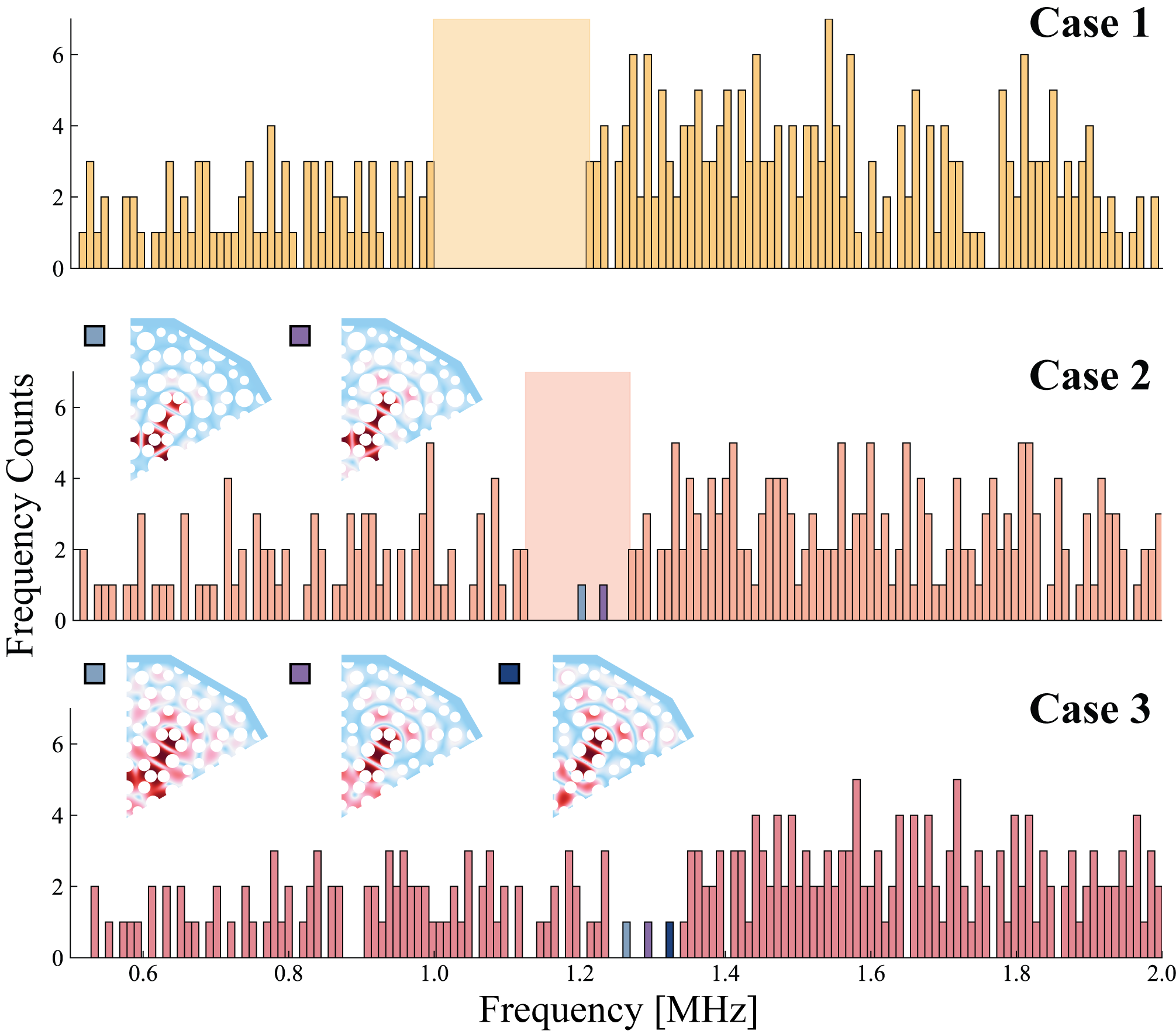}
    \caption{\textbf{Data-driven identification of stopbands for three representative design cases.} Shaded regions denote the identified stopbands. Insets show representative mode shapes within these frequency intervals, highlighting clean stopbands (Case~1), isolated localized defect modes (Case~2), and modes with boundary leakage indicating a poorly formed stopband (Case~3).}
    \label{fig:bandgap_validation}
\end{figure}

\begin{figure*}[t]
\includegraphics[width=1\linewidth]{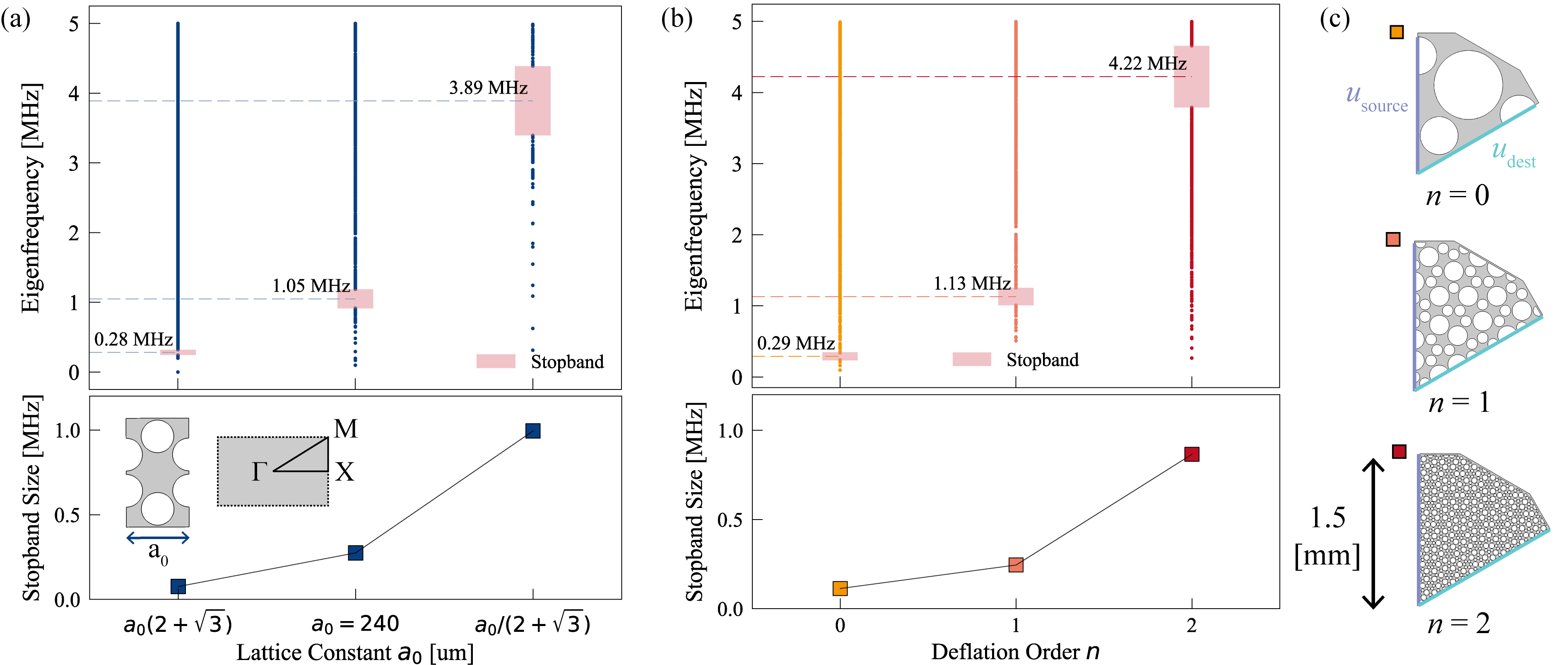}
\caption{\label{fig:bandgap_result}
\textbf{Comparison of stopband behavior between periodic phononic crystal and quasicrystal nanomechanical resonator designs for Design~1.}
(a) Stopband evolution in phononic crystal-based designs as a function of lattice constant $a_0$ ($r = 0.26a_0$). 
(b) Stopband evolution in quasicrystal-based designs as a function of deflation order $n$. 
(c) Representative quasicrystal geometries for $n=0,1,2$, illustrating the geometric refinement and the symmetry-reduced sector with cyclic boundary conditions.
}
\end{figure*}

The formation of well-defined stopbands is a prerequisite for realizing higher-order soft-clamped modes in QC-based nanomechanical resonators.
Therefore, we first establish that quasi-periodic architectures support systematically identifiable stopbands, prior to optimizing their mechanical quality factors $Q_m$.
In the following, we consider two representative QC-based resonator architectures, Design~1 and Design~2, which differ in their mass-contrast strategies and structural motifs, as defined in Method Section~\ref{sec:4.A}.
Fig.~\ref{fig:bandgap_validation} presents representative stopband identification results for three parametric variants based on Design~1, varying the geometric parameters while keeping the quasicrystal topology fixed.
Case~1 exhibits a clean and wide stopband without defect modes. 
Cases~2 and~3 illustrate situations in which isolated modes appear within the candidate stopband, making classification based solely on the eigenfrequency spectrum ambiguous.
The proposed approach consistently distinguishes between well-formed stopbands containing isolated defect modes and regions of sparse eigenmodes arising from poorly formed stopbands.
The corresponding mode shapes corroborate this distinction: in Case~2, the modes are strongly localized near the center of the resonator, whereas in Case~3, pronounced leakage toward the clamping boundary is observed.
Here, localized defect modes are introduced at the resonator center, while the surrounding QC geometry governs the formation and quality of the supporting stopband.

Notably, the identified stopbands consistently separate spatially localized defect modes from extended modes that couple strongly to the boundary, providing a clear physical distinction between functional and non-functional resonance states. 
This separation forms the basis for selectively targeting defect modes favorable for high-$Q_m$ operation.
The stopband identification remains qualitatively robust across a wide range of geometric parameters and clustering settings (see Appendix \ref{appendix:bandgap_identification} and Fig.~\ref{fig:appendix_dbscan_validation}).
Similar behavior is observed for QCs with different fold symmetries (see Appendix~\ref{appendix:alt_qc_structure}, Fig.~\ref{fig:appendix_alt_qc_structures}, and Table~\ref{tab:appendix_alt_qc_structure}), indicating that the robustness arises from the underlying quasi-periodic order rather than a specific geometric realization.
An additional frequency-response analysis of Design~2 further supports the identification of this frequency range as a stopband: within the identified stopband, the off-defect responses are strongly suppressed, while a pronounced resonance remains localized at the central defect near $f/f_0=1$ (see Appendix~B.4 and Fig.~B.4).

A direct comparison between stopband behavior in periodic PnC-based and QC-based resonator designs is shown in Fig.~\ref{fig:bandgap_result}, using Design~1 as a representative example. 
In periodic PnC structures, stopband frequencies follow Bloch scaling and vary inversely with the lattice constant $a_0$~\cite{kushwaha_theory_1994}, as confirmed by uniformly scaling the design in Fig.~\ref{fig:bandgap_result}(a). 
The stopband width also increases as $a_0$ is reduced. 
Analytically, the stopband size scales approximately as $l_t^{-1} w_t^{-1/2}$, where $l_t$ and $w_t$ denote the effective length and width of the thin interconnecting elements~\cite{reetz_analysis_2019}. 
When the geometric ratio between masses and connectors is maintained, both $l_t$ and $w_t$ decrease linearly with $a_0$, resulting in the observed nonlinear increase in stopband width.

In contrast, the stopband frequency and width in QC-based resonators are governed by the deflation order $n$ rather than a single lattice constant, as shown in Fig.~\ref{fig:bandgap_result}(b). 
Each increment in $n$ introduces a geometric scaling factor of $1/(2+\sqrt{3})$, refining the structure and shortening its characteristic length scales, which leads to a systematic upward shift in stopband frequency (Fig.~\ref{fig:bandgap_result}(c)). 
Unlike periodic PnCs, analytical estimates of the stopband width are not readily available for the QC architectures considered here. Within the deflation orders investigated, our simulations show a consistent increase in the stopband frequency and width. However, the present data-driven study does not provide a general proof of monotonic broadening at arbitrarily large deflation orders. Establishing whether this trend persists at higher deflation orders would require further development of a physics-based theoretical framework for QC architectures. A similar broadening trend is also reproduced in Design~2 (see Appendix~\ref{appendix:bandgap_identification}, Fig.~\ref{fig:appendix_bandgap_HMC}, and Method Section~\ref{sec:4.A}), supporting the robustness of the observed trend across the two substantially different design strategies investigated here.
These findings demonstrate that well-defined stopbands emerge in QC architectures through quasi-periodic scaling mechanisms. Although distinct from Bloch-type periodic systems, they remain systematically accessible for resonator design through our data-driven approach.

\subsection{\label{sec:2.B} Soft-Clamped Defect Modes in Quasicrystal Architectures}

\begin{figure*}[t]
\centering
\includegraphics[width=\linewidth]{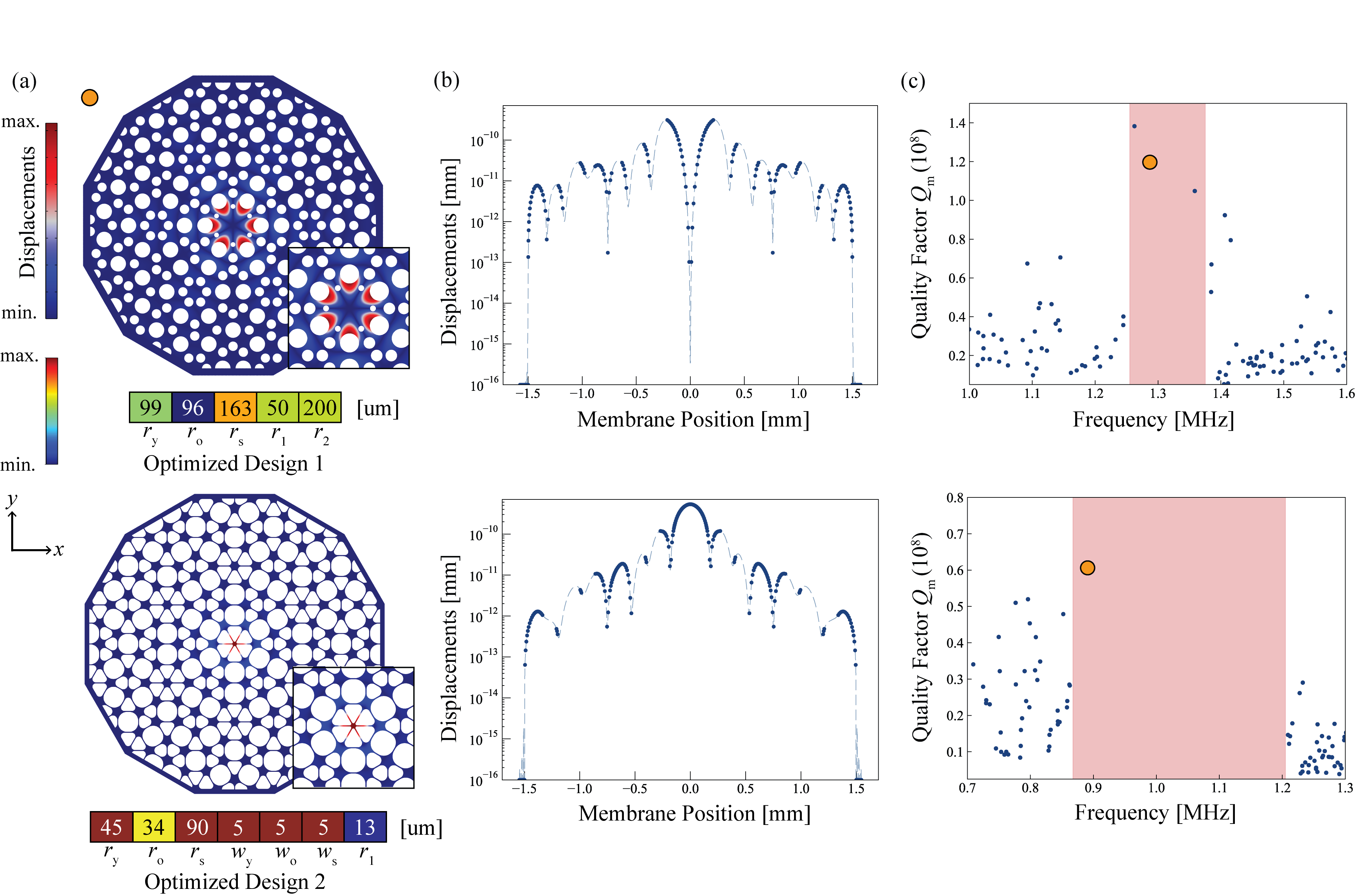}
\caption{
\textbf{Optimized Designs 1 and 2.}
(a) Optimized mode shapes for the representative Designs 1 and 2 (see Fig.~\ref{fig:design_formulation_method} for detailed design formulations), with insets highlighting the localized defect modes and the corresponding optimized design parameters.
(b) Displacement profiles along the $x$-axis through the membrane center, showing strong spatial confinement and suppression of motion toward the clamping boundary.
(c) Simulated mechanical quality factors $Q_m$ for all modes within the identified stopband, with highlighted markers indicating the selected defect modes.
}
\label{fig:result_mode_shape}
\end{figure*}

Having established robust stopband identification in QC-based resonators, we now exploit them to realize soft-clamped modes with enhanced mechanical performance. 
Modes residing within a clean stopband are strongly localized near the defect region and decoupled from the clamping boundary, consistent with periodic PnC resonators~\cite{tsaturyan_ultracoherent_2017}. 
As illustrated in Fig.~\ref{fig:result_mode_shape}, this spatial confinement suppresses bending-related dissipation by attenuating vibrational motion toward the supports, forming the physical basis of soft clamping in QC-based resonators.
The optimized designs shown in Fig.~\ref{fig:result_mode_shape} are obtained by tuning a compact set of QC tile parameters, including string widths and hole or fillet radii (Method Section~\ref{sec:4.A}).

Within each identified stopband, candidate defect modes are selected based on their spatial localization and minimal displacement at the clamping boundary, and their mechanical quality factors are then evaluated using the energy-based proxy (Method Section~\ref{sec:4.B}). 
Strong spatial confinement suppresses bending-induced dissipation, enabling high $Q_m$ values across distinct mass regimes.
Table~\ref{tab:design_performance_table} summarizes representative metrics for two optimized designs, where $Q_m$ is taken as the primary optimization target.
The $Q_m$ values reported here are numerically predicted and account only for intrinsic material loss as modeled by dissipation dilution. Although previous studies of soft-clamped nanomechanical resonators have generally reported good agreement between numerical predictions and experimental measurements~\cite{shin_spiderweb_2022,cupertino_centimeter-scale_2024,tsaturyan_ultracoherent_2017,ghadimi_radiation_2017}, direct experimental validation remains necessary for the proposed QC architectures. Additional loss channels, including imperfect clamping, fabrication-induced defects, and geometric deviations, may reduce the experimentally measured $Q_m$ values.
For quantum probing, the thermal coherence time $\tau$ must exceed at least one oscillation period ($1/f_m$)~\cite{marquardt_quantum_2007,marshall_towards_2003,tsaturyan_ultracoherent_2017}. 
At room temperature ($T=300$ K), this condition reduces to $Q_m f_m > 6 \times 10^{12}$ Hz, which is satisfied by all optimized designs. 
Designs~1 and~2 represent low- and high–mass-contrast implementations of the same QC framework, confirming the robustness of stopband-enabled soft clamping across distinct structural motifs.

\begin{table}[b]
\caption{\label{tab:design_performance_table}Numerically predicted performance of optimized Designs~1 and~2, shown in Fig.~\ref{fig:result_mode_shape} with the mechanical quality factor $Q_m$ as the primary optimization target, assuming room temperature ($T = 300\,\mathrm{K}$).}
\begin{tabular}{l|c|c}
\textrm{Designs} & \textrm{1} & \textrm{2} \\
\hline
Frequency $f$ (MHz) & 1.29 & 0.88 \\
Effective mass $m_\mathrm{eff}$ (ng) & 5.9 & 0.92 \\
Dilution factor $D_Q$ ($\times 10^3$) & 48.9 & 25.1 \\ 
Quality factor $Q_m$ ($\times 10^6$) & 118.3 & 60.5 \\
$Qf$ product (THz) & 153 & 53.3 \\
Coherence time $\tau$ ($\mu s$) & 3.0 & 1.54 \\
$\sqrt{S_{FF}}$ (aN/$\sqrt{\mathrm{Hz}}$) & 58.0 & 26.4 \\
\end{tabular}
\end{table}

\begin{figure*}[t]
\centering
\includegraphics[width=\linewidth]{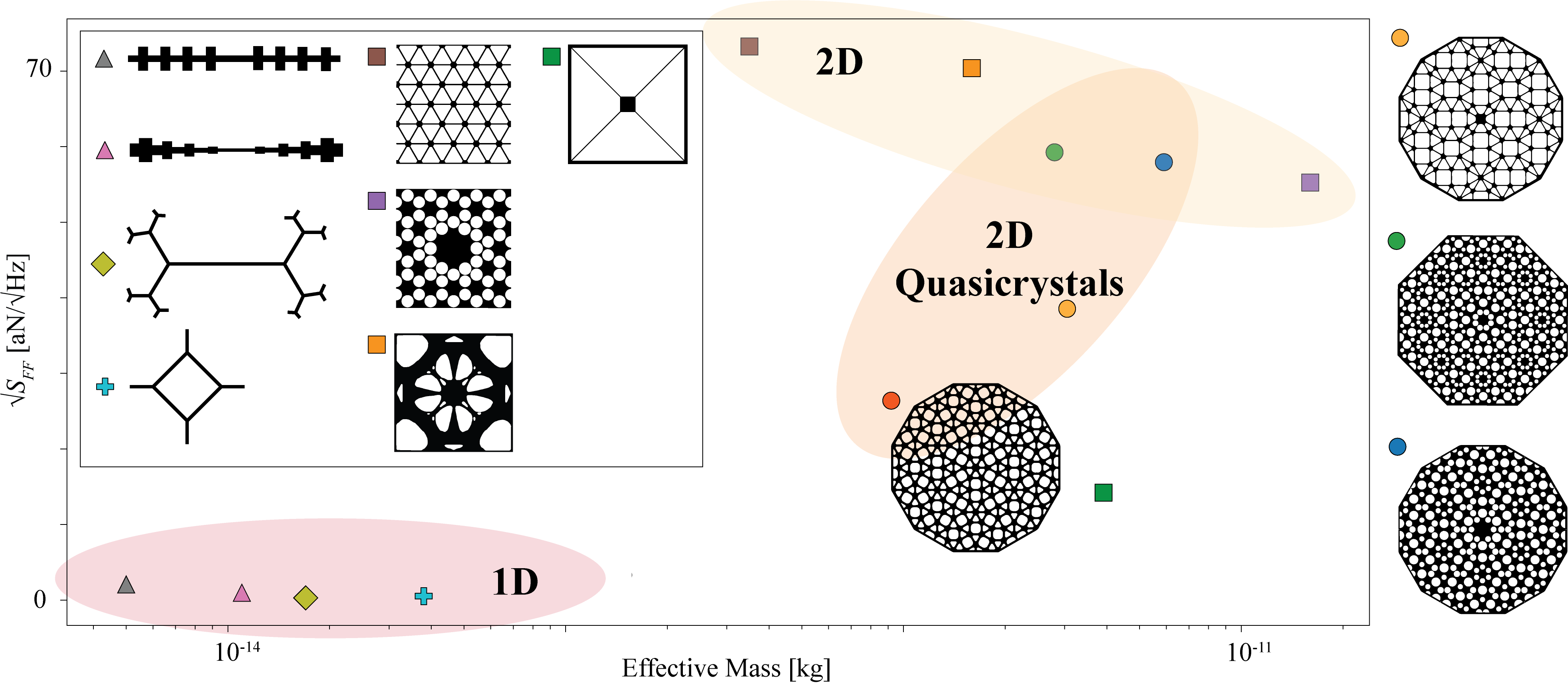}
\caption{\textbf{Thermal force noise--limited sensitivity ($\sqrt{S_{FF}}$) versus effective mass for representative nanomechanical resonator architectures.}
One-dimensional string-like resonators achieve exceptional sensitivity through extremely low effective mass, whereas two-dimensional pad-based resonators offer improved optical accessibility at the cost of increased mass.
The QC-based designs presented in this work begin to occupy an intermediate regime, approaching the force sensitivity of one-dimensional resonators while retaining a fundamentally two-dimensional geometry.
The orange circle denotes the best-performing design, while other circles indicate additional QC realizations (see Appendix~\ref{appendix:alt_qc_structure} and Fig.~\ref{fig:appendix_alt_qc_structures}).}
\label{fig:ashby_plot}
\end{figure*}

The performance trade-off between effective mass and mechanical dissipation can be quantified through the thermally limited force noise~\cite{saulson_thermal_1990}, given by $S_{FF} = 4\pi m_{\mathrm{eff}} f_m k_B T / Q_m.$
Owing to the strong spatial localization enabled by QC stopbands, the optimized Design~2 predicts an effective mass of $m_{\mathrm{eff}} = 0.92$~ng, resulting in a thermal force noise–limited sensitivity of $\sqrt{S_{FF}} = 26.4$~aN$/\sqrt{\mathrm{Hz}}$. 
To contextualize this performance, we compare the achieved sensitivity and effective mass against representative one- and two-dimensional nanomechanical resonator architectures in Fig.~\ref{fig:ashby_plot}. 
A direct simulation-based comparison with a representative periodic PnC resonator under comparable material and geometric conditions is provided in Appendix~\ref{appendix:benchmark_pnc}.
One-dimensional string-like PnC resonators~\cite{ghadimi_elastic_2018} achieve exceptional sensitivities through extremely low effective mass and strong clamping-loss suppression, but lack a central pad for efficient free-space optical coupling. 
Conversely, two-dimensional membrane- and pad-based PnC resonators provide improved optical accessibility~\cite{aspelmeyer_cavity_2014,enzian_phononically_2023,zhou_cavity_2023,chen_high-finesse_2017,norte_mechanical_2016,shi_topology_2026,reinhardt_ultralow_noise_2016}, typically at the cost of increased effective mass and reduced sensitivity. 
The QC-based designs presented here operate in an intermediate regime between these paradigms: by leveraging QC stopbands within a fundamentally two-dimensional geometry, they approach the force sensitivity of one-dimensional resonators while retaining the structural and optical advantages of pad-based architectures. Although the limited availability of consistently reported performance metrics in the literature restricts the number of directly comparable reference structures, the comparison suggests that QC architectures provide a viable alternative to periodic PnC designs, potentially opening a previously underexplored design space that combines the force sensitivity of one-dimensional resonators with the structural and optical advantages of two-dimensional pad-based systems.

\section{\label{sec:3} Discussion}
In this work, we introduced and validated a new class of nanomechanical resonator architectures based on quasicrystal (QC) geometries.
By leveraging quasi-periodic order, we demonstrated that well-defined mechanical stopbands and strongly localized defect modes can be systematically identified and exploited despite the absence of translational periodicity.
Rather than merely adapting existing soft-clamping concepts, QC architectures fundamentally relax the requirement of translational periodicity, revealing that dissipation dilution can emerge from quasi-periodic order itself.
Collectively, these results demonstrate that translational periodicity is not a fundamental prerequisite for stopband-enabled dissipation dilution, and that quasi-periodic order can serve as a scalable structural prior for soft-clamped nanomechanical resonators.

Using a representative 12-fold QC design, we numerically predicted a thermal force noise–limited sensitivity of $26.4$~aN$/\sqrt{\mathrm{Hz}}$, placing its performance between state-of-the-art one-dimensional string-like devices and two-dimensional pad-based membrane designs.
Beyond this specific realization, multiple QC geometries with different rotational symmetries exhibited consistently strong baseline performance, highlighting the generality of QC architectures as a viable structural platform for high-performance nanomechanical resonators.
Without translational symmetry, analytic band-structure methods break down, making data-driven spectral inference indispensable for QC design.
Although the present conclusions are drawn from simulation-driven optimization, the observed trends—robust mode localization, stopband-enabled suppression of clamping loss, and consistent cross-design performance—are expected to persist experimentally, even if absolute $Q_m$ values may vary with fabrication-specific loss channels~\cite{shin_spiderweb_2022,cupertino_centimeter-scale_2024}.

While this study focuses on optimization-driven identification and exploitation of QC stopbands for high-performance soft clamping, QC-based resonator architectures exhibit additional features that arise from their aperiodic geometry. 
Unlike periodic lattices defined by a single unit cell and lattice constant, QC structures embed hierarchical length scales and symmetry-controlled local environments, enabling multiscale control over spectral and localization characteristics. 
In particular, the deflation order provides a direct geometric knob for tuning the stopband frequency scale, offering a systematic route to position soft-clamped modes within target frequency windows. 
Moreover, the structural tendency of QC patterns toward the center facilitates integration with in-plane strain engineering, allowing stress localization near the defect region without fundamentally altering stopband-supported confinement (see Appendices~\ref{appendix:other_studies_inplane_se} and \ref{appendix:other_studies_device_size}, and Figs~\ref{fig:in-plane_SE} and \ref{fig:add_layer}).
Preliminary analyses also indicate reduced sensitivity of resonance frequency to fabrication-induced geometric variations compared to representative periodic PnC designs, with systematic parametric studies provided in Appendix~\ref{appendix:other_studies_robustness} and Fig.~\ref{fig:appendix_robustness}.
A comprehensive analytical theory of QC stopband scaling remains an open challenge and may offer deeper insight into the spectral trends observed here. 
These numerical observations suggest that QC architectures may have practical relevance beyond demonstrating theoretical feasibility. The reduced sensitivity to geometric variations observed in the representative comparison, together with their compatibility with strain engineering and size-dependent stress localization, motivates further development toward experimentally realizable QC resonators. Although fabrication yield, process tolerance, and manufacturing cost have not been assessed in the present study, they remain important topics for future experimental investigation.

Looking forward, the combination of QC geometries with data-driven design tools expands the resonator design space beyond periodic architectures. 
The aperiodic nature of QC introduces structural degrees of freedom, including hierarchical scaling and intrinsic mode localization, that can be systematically explored within a unified framework.
However, the QC architectures explored here represent only a small branch of the much broader QC family; ongoing exploration and comparison across QC architectures will further establish QC-based designs as a viable platform for nanomechanical resonators.
In this sense, the present approach is complementary to inverse-design techniques such as topology optimization~\cite{shi_topology_2026,algra_dissipation_2025,gao_systematic_2020,hoj_ultracoherent_2024,hoj_ultra-coherent_2021}.
Whereas topology optimization is effective for refining local defect geometries, QC architectures provide background structures that inherently support higher-order stopbands and robust soft clamping, embedding localization at the structural level as a scalable geometric prior for further optimization.

Finally, beyond structural considerations, the intrinsic phononic physics of QC architectures presents additional opportunities.
Quasi-periodic order has been shown to give rise to unconventional vibrational transport phenomena, including symmetry-breaking propagation effects~\cite{matsuura_singular_2024}.
Such behavior may be exploited to suppress directional energy leakage or engineer enhanced phononic isolation in future resonator designs.
More broadly, the interplay between quasi-periodic order and phononic transport may enable anisotropic mode confinement and disorder-tolerant localization that are difficult to realize in strictly periodic systems.
Taken together, these perspectives position quasicrystal architectures not merely as geometric variants of phononic crystals, but as a broader structural paradigm for engineering dissipation, localization, and wave isolation in nanomechanical systems.

\section{\label{sec:4} Method}
\subsection{\label{sec:4.A} Geometric Parameterization of Quasicrystal Resonators}

\begin{figure*}[t]
\includegraphics[width=\linewidth]{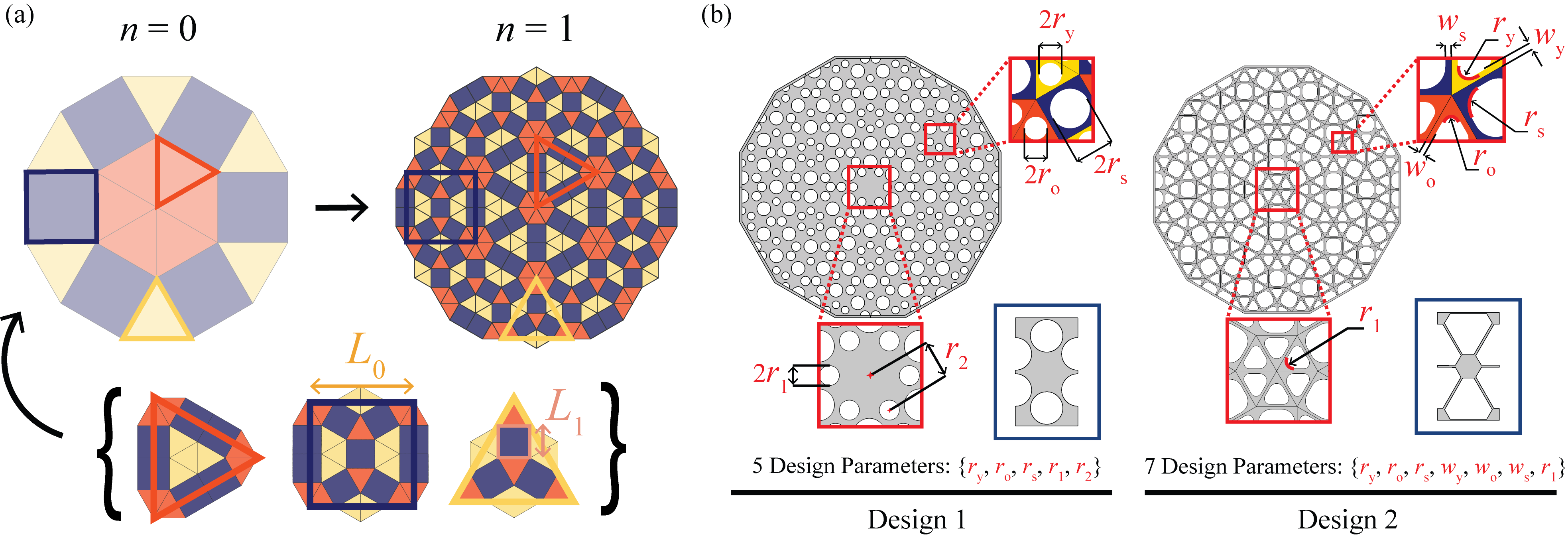}
\caption{\label{fig:design_formulation_method}
\textbf{Design formulation of 12-fold quasicrystal-based nanomechanical resonators.
}(a) Stampfli deflation process for generating a 12-fold dodecagonal quasicrystal lattice, illustrating the transition from the zeroth-order ($n=0$) to the first-order ($n=1$) structure and the associated characteristic length scales $L_0$ and $L_1$.
(b) Parameterized resonator designs implemented on the quasicrystal lattice using two representative phononic crystal–inspired motifs.
Design~1 employs a low mass-contrast strategy based on hole patterns, while Design~2 adopts a high mass-contrast strategy using massive pads connected by thin strings.
The geometric design parameters, including string widths ($w_i$) and hole or fillet radii ($r_i$), are defined with respect to individual quasicrystal tile types.
}
\end{figure*}

Since the seminal work of Levine and Steinhardt, which established quasicrystal (QC) as a distinct class of ordered structures~\cite{levine_quasicrystals_1984,levine_quasicrystals_1986}, a wide variety of quasicrystalline geometries have been investigated across photonic, phononic, and mechanical systems.
Motivated by prior studies demonstrating that 12-fold QC can support stopband characteristics comparable to those of two-dimensional photonic crystals~\cite{zoorob_complete_2000}, we focus on a representative 12-fold dodecagonal QC architecture as a reference platform.
In the following, we introduce the geometric construction of the QC lattice and the parameterized resonator formulation used throughout this study.
Extensions to QCs with other rotational symmetries are summarized in Appendix~\ref{appendix:alt_qc_structure} and Fig.~\ref{fig:appendix_alt_qc_structures}, demonstrating that the proposed framework is symmetry-agnostic and extends beyond the specific 12-fold geometry considered here.

The 12-fold dodecagonal QC structure is constructed following the Stampfli deflation rule, as illustrated in Fig.~\ref{fig:design_formulation_method}(a). 
We begin with an initial tiling composed of three distinct tile types, corresponding to a square and two types of triangles.
Each tile is subsequently replaced by a corresponding substitution pattern containing finer geometric features, resulting in a transition from a zeroth-order to a first-order deflated QC structure. 
The deflation order, denoted by $n$, directly controls the structural density of the QC. 
For the 12-fold dodecagonal QC, successive deflation follows the scaling relation 
\begin{equation}
    L_{n+1} = \frac{L_n}{2+\sqrt{3}},
    \label{eqn:deflation_relation}
\end{equation}
where $L_n$ is the characteristic lattice length at the $n^{\rm th}$ deflation level. 
In principle, the deflation procedure can be iterated to higher orders, limited primarily by computational constraints (see Appendix~\ref{appendix:12_fold_geometry} and Fig.~\ref{fig:appendix_deflation_rule}).
This hierarchical construction introduces a well-defined and controllable length scale, which enables systematic exploration of the resonator design space.

Having established the geometric construction of the QC lattice, we now describe how this hierarchical geometry is translated into concrete nanomechanical resonator designs.
As illustrated in Fig.~\ref{fig:introduction}, a given QC geometry can be combined with different structural design motifs, leading to distinct nanomechanical resonator architectures.
Rather than imposing additional geometric assumptions, the present formulation directly exploits a defining feature of quasicrystals—the existence of distinct tile types—such that the resonator geometry is parameterized in a manner that is naturally aligned with the underlying QC tiling.

To assess the feasibility and performance of QC-based nanomechanical resonators, two representative design motifs inspired by membrane-based phononic crystal (PnC) resonators are considered.
These motifs correspond to low and high mass-contrast strategies that have proven effective in realizing stopband-enabled soft clamping in periodic PnC structures~\cite{tsaturyan_ultracoherent_2017,reetz_analysis_2019}.
In the present work, both motifs are implemented on first-order QC geometries to enable direct benchmarking against established PnC-based designs, as summarized in Fig.~\ref{fig:design_formulation_method}(b).

Building on the intrinsic tiling patterns of the QC lattice, the resonator geometry is parameterized using a compact set of geometric variables associated with individual tile types.
Specifically, the widths of interconnecting strings ($w_i$) and the radii of holes or fillets ($r_i$) are used as design variables, resulting in a low-dimensional yet expressive design space suitable for quantitative comparison and optimization.
Localized defect regions are introduced to support soft-clamped modes by forming a central pad.
While more elaborate defect designs can significantly influence resonator performance~\cite{tsaturyan_ultracoherent_2017,halg_membrane-based_2021,engelsen_ultrahigh-quality-factor_2024}, the simplified defect implementations adopted here provide a controlled baseline for evaluating QC-enabled soft clamping.
All designs are compatible with large-area fabrication using standard photolithography and are based on high-stress $\text{Si}_3\text{N}_4$ membranes~\cite{shin_spiderweb_2022}, a well-established platform for high-$Q_m$ nanomechanical resonators.
Further details regarding parameter bounds, fabrication constraints, and material properties are provided in Appendix~\ref{appendix:design_parameters}.

\subsection{\label{sec:4.B} Data-Driven Design of High-$Q_m$ Quasicrystal Resonators}

Given the high fabrication cost and limited throughput of high–aspect ratio nanomechanical membranes~\cite{norder_pentagonal_2025}, the design of soft-clamped resonators increasingly relies on numerical simulations. Finite-element analysis has been shown to provide reliable predictions of mode shapes, frequencies, and dissipation trends once the soft-clamping regime is established~\cite{shin_spiderweb_2022,cupertino_centimeter-scale_2024}. The present study is formulated within a simulation-driven design framework.
However, extending established simulation workflows from periodic PnC to QC architectures introduces a fundamental limitation: the absence of long-range translational symmetry precludes unit-cell-based band-structure analysis, which underlies conventional stopband identification. 
As a result, both the identification and exploitation of stopbands in QCs require a qualitatively different, data-driven formulation.

In conventional PnC resonators, stopbands are typically identified through unit-cell band-structure analysis, after which the whole device is analyzed only within the relevant frequency windows. 
This approach is computationally efficient but relies on translational periodicity, which makes it unsuitable for QC architectures. 
Despite the absence of translational symmetry, QCs retain well-defined rotational order. 
In the 12-fold dodecagonal QC studied here, this discrete symmetry is used as a spectral sampling mechanism to efficiently probe the vibrational spectrum, without invoking Bloch periodicity or periodic approximants.
Adjacent $30^\circ$ sectors are related by reflection, whereas the smallest domain that repeats by pure rotation spans $60^\circ$, corresponding to a 6-cyclic symmetry.
Specifically, eigenfrequency analysis is performed on a symmetry-reduced sector corresponding to one-sixth of the whole structure, using cyclic Floquet-type boundary conditions that relate displacement fields on opposing edges through a combined phase shift and rotation,
\begin{eqnarray}
    \vec{u}_{\text{dest}} = \text{exp}(i\varphi) \mathbf{R}_{\theta} \vec{u}_{\text{source}},
    \label{eqn:cyclic_periodic_bc}
\end{eqnarray}
where $\mathbf{R}_{\theta}$ is the rotation matrix for $\theta = \pi/3$ and $\varphi = m\theta$ with integer $m$. 
For this 6-cyclic decomposition, the phase sectors are $m=0,\ldots,5$. Since the underlying structural eigenvalue problem is real-valued, the $m$ and $6-m$ sectors form complex-conjugate pairs with identical eigenfrequencies. Therefore, $m \in \{0,1,2,3\}$ is sufficient to sample the distinct eigenfrequency branches.
The analysis proceeds in two stages: a stationary prestress step that captures stress redistribution, followed by a linearized eigenfrequency analysis about the prestressed state. 
This symmetry-reduced procedure produces a dense one-dimensional spectrum of eigenfrequencies capturing the relevant spectral features of the full QC resonator, while reducing computational cost by more than an order of magnitude. 
Importantly, it provides a physically consistent foundation for the subsequent identification of stopbands.
Furthermore, the symmetry-reduced eigenfrequency results are compared with a full-structure simulation in Appendix~\ref{appendix:stopband_validation}, where both models show consistent spectral trends and identify the same stopband region.

Based on the resulting eigenfrequency spectrum, stopbands are identified as frequency intervals in which the density of eigenmodes is strongly suppressed. 
To detect such intervals systematically, we employ an unsupervised density-based clustering approach. 
Stopbands are identified using the density-based clustering algorithm DBSCAN~\cite{yin_rapid_2024}, with $MinPts=5$ and a dynamically tuned $Eps$ (see Appendix~\ref{appendix:bandgap_identification_parameter} and Fig.~\ref{fig:appendix_dbscan}). 
Stopbands with widths narrower than $100$~kHz are excluded from further analysis.

Subsequently, a second prestressed eigenfrequency analysis is performed on the full structure, restricted to the identified stopband frequency windows, in order to resolve localized defect modes in detail. The mechanical quality factor $Q_m$ of each design is estimated using an energy-based proxy defined as the ratio between the maximum kinetic energy $E^{\mathrm{max}}_{\mathrm{kin}}$ and the energy dissipated per cycle due to bending loss $W_{\mathrm{bend}}$,
\begin{equation}
Q_m = 2 \pi\frac{ E^{\mathrm{max}}_{\mathrm{kin}}}{W_{\mathrm{bend}}}= D_Q Q_0.
\label{eqn:Q_calculation}
\end{equation}
Here, $D_Q$ denotes the dissipation dilution factor. The intrinsic quality factor $Q_0$ of the $\mathrm{Si}_3\mathrm{N}_4$ membrane is expressed as $Q_0^{-1} = Q_{\mathrm{vol}}^{-1} + Q_{\mathrm{surf}}^{-1}$, where $Q_{\mathrm{vol}}$ represents bulk material loss and $Q_{\mathrm{surf}}$ denotes surface loss, which depends linearly on the membrane thickness $t$~\cite{villanueva_evidence_2014}. For thin membranes ($t \lesssim 100$~nm), surface loss dominates, and we adopt the empirical approximation $Q_0 \approx 6900\,t/100~\mathrm{nm}$~\cite{ghadimi_elastic_2018}. The maximum kinetic energy and the energy dissipated per cycle due to bending loss are defined as:
\begin{align*}
E^{\text{max}}_{\text{kin}}
&= \frac{\rho t}{2}\iint \omega_m^2 u^2\,dx\,dy, \\
W_{\text{bend}}
&= \frac{\pi Et^3}{12Q_0(1-\nu^2)}
   \iint \Bigl[ u_{xx}^2 + u_{yy}^2 \\
&\qquad + 2\nu u_{xx}u_{yy}
   + 2(1-\nu)u_{xy}^2 \Bigr]\,dx\,dy,
\end{align*}
where $u$ denotes the out-of-plane displacement field, subscripts indicate partial derivatives with respect to spatial coordinates, and $\omega_m$ is the angular resonance frequency.

Because $Q_m$ depends on higher-order displacement gradients, fine-mesh refinement is required in regions of concentrated bending, such as joints and boundaries, leading to a substantial increase in computational cost. 
A detailed description of the simulation setup with mesh control can be found in Appendix~\ref{appendix:simulation} and Fig.s~\ref{fig:appendix_simulation_pipeline}, \ref{fig:appendix_massive_LMC_mesh} and \ref{fig:appendix_massive_HMC_mesh}.
As a result, each design evaluation becomes significantly more expensive, rendering exhaustive parameter sweeps impractical. 
To efficiently explore the design space under these constraints, we employ a single-objective Bayesian optimization strategy with $Q_m$ as the objective function~\cite{pelikan_boa_1999}, which is seen effective for material discovery problems involving similarly expensive experiments~\cite{schmidt_2019_recent,lookman_2019_active} and is well suited to the present data-scarce, high-cost design setting~\cite{shin_spiderweb_2022,cupertino_centimeter-scale_2024}. Details on the optimization algorithm can be found in Appendix~\ref{appendix:bayesian_optimization} and Fig.~\ref{fig:appendix_bayesian_opt}.

\section*{Data Availability Statement}
The data that support the findings of this study are available from the corresponding authors upon reasonable request.

\section*{Conflict of Interest}
The authors declare no conflict of interest.

\section*{CRediT Author Contributions}

\textbf{Kawen Li:} Conceptualization, Data curation, Formal analysis, Methodology, Investigation, Validation, Visualization, Writing - original draft, Writing - review and editing. 

\textbf{Hangjin Cho:} Data curation, Investigation, Validation, Additional experiments, Visualization, Writing - review and editing

\textbf{Richard Norte:} Conceptualization, Funding acquisition, Investigation, Project administration, Resources, Software, Supervision, Writing - review and editing

\textbf{Dongil Shin:}
Conceptualization, Funding acquisition, Investigation, Project administration, Resources, Software, Supervision, Writing - review and editing

\clearpage
\onecolumngrid
\appendix
\renewcommand{\thesubsection}{\thesection.\arabic{subsection}}
\renewcommand{\theHfigure}{\thesection.\arabic{figure}}
\renewcommand{\theHequation}{\thesection.\arabic{equation}}
\renewcommand{\theHtable}{\thesection.\arabic{table}}

\section{Deflation Rules for 12-Fold Dodecagonal Quasicrystals}
\label{appendix:12_fold_geometry}

\setcounter{figure}{0}
\setcounter{equation}{0}
\setcounter{table}{0}
\renewcommand{\thefigure}{A.\arabic{figure}}
\renewcommand{\theequation}{A.\arabic{equation}}
\renewcommand{\thetable}{A.\arabic{table}}

\begin{figure}[h]
    \centering
    \includegraphics[width=\linewidth]{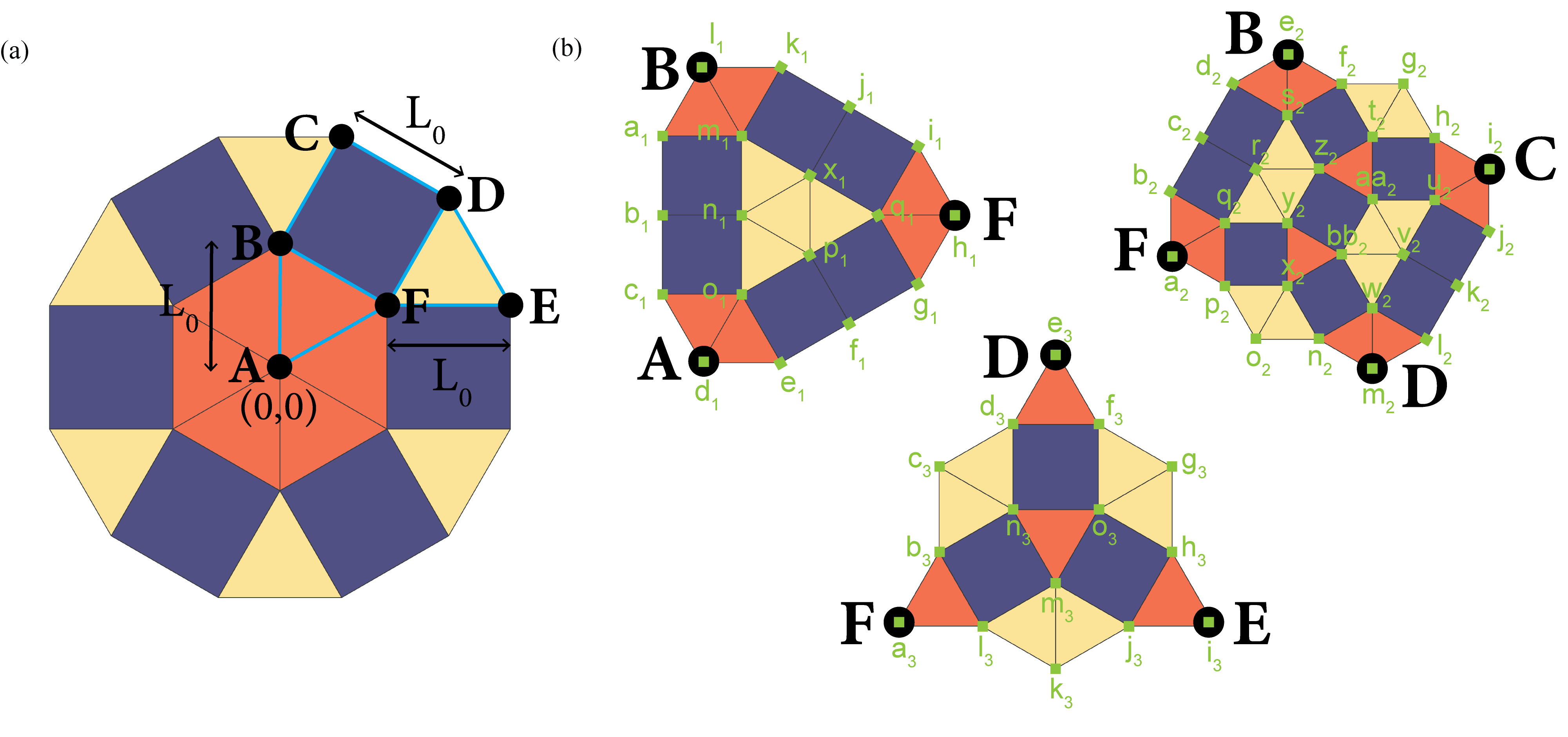}
    \caption{\textbf{Detailed deflation rules for 12-fold dodecagonal quasicrystal lattice with stampfli deflation rules.}
    (a) Base unit cell structure ($n=0$) used for our design with $L_0 = 1.5/(1+\sqrt{3}/2)$~[mm]. Major coordinates are marked in black for a cyclic symmetric slice. (b) Sub-structures used to deflate the base unit cell structure. By matching the marked coordinates in black, one can substitute the sub-structures into the base unit cell structure, giving a first-order ($n=1$) structure.}
    \label{fig:appendix_deflation_rule}
\end{figure}

For the 12-fold dodecagonal quasicrystal (QC) structure studied in this work, the geometry is constructed from three distinct tiles: squares, orange triangles, and yellow triangles. To generate the corresponding QC structure, one begins with a base unit cell configuration (Fig.~\ref{fig:appendix_deflation_rule}(a)) and applies a substitution rule that replaces each tile with a specific pattern, as shown in Fig.~\ref{fig:appendix_deflation_rule}(b). This substitution process defines the deflation rule of the QC structure. Given the coordinates of all tiles within the base unit cell, the coordinates of the newly substituted tiles can be recursively computed.

To mathematically model the deflation process, we assign labels to the major vertices of the base unit cell using capital letters $\text{A}$ through $\text{F}$, and to the vertices of the substituted tiles using lowercase labels $\text{a}_i$ through $\text{bb}_i$, where $i$ indexes the three tile types (square, orange triangle, and yellow triangle). In this description, we restrict our derivation to the cyclic-symmetric sector of the base unit cell (highlighted in blue in Fig.~\ref{fig:appendix_deflation_rule}(a)), since the full 12-fold structure can be constructed by rotating this sector five times in increments of $\pi/3$ radians.

Assuming point $\text{A}$ lies at the origin of the global cartesian coordinate system, the positions of all other major vertices can be defined as vectors originating from the origin:

\begin{align}
\vec{A} &= a_x \hat{\imath} + a_y \hat{\jmath} = 0 \notag \\
\vec{B} &=  b_x \hat{\imath} + b_y \hat{\jmath} = L_0 \hat{\jmath} \notag \\
\vec{C} &=  c_x \hat{\imath} + c_y \hat{\jmath} = \vec{B} + \left(\frac{1}{2} \hat{\imath} + \frac{\sqrt{3}}{2} \hat{\jmath} \right) L_0 \notag \\
\vec{D} &=  d_x \hat{\imath} + d_y \hat{\jmath} = \vec{F} + \left(\frac{1}{2} \hat{\imath} + \frac{\sqrt{3}}{2} \hat{\jmath} \right) L_0 \notag \\
\vec{E} &=  e_x \hat{\imath} + e_y \hat{\jmath} = \vec{F} + L_0 \hat{\imath} \notag \\
\vec{F} &=  f_x \hat{\imath} + f_y \hat{\jmath} = \left(\frac{\sqrt{3}}{2} \hat{\imath} + \frac{1}{2} \hat{\jmath} \right) L_0
\label{eqn:appendix_major_coordinates}
\end{align}

To facilitate the computation of the substituted tile coordinates, we define a set of directional unit vectors. These vectors serve as local references for transformations such as translations and rotations within the deflation rule. As an illustrative example, we show the directional unit vector from point $\text{A} = a_x\hat{\imath} + a_y \hat{\jmath}$ to $\text{B} = b_x\hat{\imath} + b_y \hat{\jmath}$ and its corresponding orthogonal vector:
\begin{align}
\vec{u}_{AB} &= \frac{\vec{B} - \vec{A}}{\|\vec{B} - \vec{A}\|}, \notag \\
\vec{u}_{AB}^\perp &= \frac{(a_y - b_y)\hat{\imath} + (b_x - a_x)\hat{\jmath}}{\sqrt{(a_y - b_y)^2 + (b_x - a_x)^2}}.
\label{eqn:appendix_directional_vector}
\end{align}

The coordinates of the vertices of the substituted orange triangle tile can then be calculated using these unit vectors. Given $L_1 = (2 - \sqrt{3})L_0$, the positions are defined as follows. 

For orange triangular tiles:
\begin{align}
\vec{a}_1 &= \vec{A} + \left(2 + \frac{\sqrt{3}}{2}\right)L_1 \vec{u}_{AB} + \frac{L_1}{2} \vec{u}_{AB}^\perp, \notag \\
\vec{b}_1 &= \vec{a}_1 - L_1 \vec{u}_{AB}, \notag \\
\vec{c}_1 &= \vec{a}_1 - 2L_1 \vec{u}_{AB}, \notag \\
\vec{d}_1 &= \vec{A}, \notag \\
\vec{e}_1 &= \vec{f}_1 + L_1 \vec{u}_{FA}, \notag \\
\vec{f}_1 &= \vec{g}_1 + L_1 \vec{u}_{FA}, \notag \\
\vec{g}_1 &= \vec{F} + \frac{\sqrt{3}}{2}L_1 \vec{u}_{FA} + \frac{L_1}{2} \vec{u}_{FA}^\perp, \notag \\
\vec{h}_1 &= \vec{F}, \notag \\
\vec{i}_1 &= \vec{j}_1 + L_1 \vec{u}_{BF}, \notag \\
\vec{j}_1 &= \vec{k}_1 + L_1 \vec{u}_{BF}, \notag \\
\vec{k}_1 &= \vec{l}_1 + \frac{\sqrt{3}}{2}L_1 \vec{u}_{BF} + \frac{L_1}{2} \vec{u}_{BF}^\perp, \notag \\
\vec{l}_1 &= \vec{B}, \notag \\
\vec{m}_1 &= \vec{a}_1 - L_1 \vec{u}_{AB}^\perp, \notag \\
\vec{n}_1 &= \vec{b}_1 - L_1 \vec{u}_{AB}^\perp, \notag \\
\vec{o}_1 &= \vec{c}_1 - L_1 \vec{u}_{AB}^\perp, \notag \\
\vec{p}_1 &= \vec{f}_1 - L_1 \vec{u}_{FA}^\perp, \notag \\
\vec{q}_1 &= \vec{g}_1 - L_1 \vec{u}_{FA}^\perp, \notag \\
\vec{x}_1 &= \vec{j}_1 - L_1 \vec{u}_{BF}^\perp.
\label{eqn:appendix_triangle_o_deflation}
\end{align}

For square tiles:
\begin{align}
\vec{a}_2 &= \vec{F}, \notag \\
\vec{b}_2 &= \vec{a}_2 + \frac{\sqrt{3}}{2}L_1 \vec{u}_{FB} + \frac{L_1}{2} \vec{u}_{FB}^\perp, \notag \\
\vec{c}_2 &= \vec{b}_2 + L_1 \vec{u}_{FB}, \notag \\
\vec{d}_2 &= \vec{c}_2 + L_1 \vec{u}_{FB}, \notag \\
\vec{e}_2 &= \vec{B}, \notag \\
\vec{f}_2 &= \vec{e}_2 + L_1 \vec{u}_{BC}, \notag \\
\vec{g}_2 &= \vec{f}_2 + \frac{\sqrt{3}}{2}L_1 \vec{u}_{BC} + \frac{L_1}{2} \vec{u}_{BC}^\perp, \notag \\
\vec{h}_2 &= \vec{f}_2 + \sqrt{3} L_1 \vec{u}_{BC}, \notag \\
\vec{i}_2 &= \vec{C}, \notag \\
\vec{j}_2 &= \vec{i}_2 + \frac{\sqrt{3}}{2}L_1 \vec{u}_{CD} + \frac{L_1}{2} \vec{u}_{CD}^\perp, \notag \\
\vec{k}_2 &= \vec{j}_2 + L_1 \vec{u}_{CD}, \notag \\
\vec{l}_2 &= \vec{k}_2 + L_1 \vec{u}_{CD}, \notag \\
\vec{m}_2 &= \vec{D}, \notag \\
\vec{n}_2 &= \vec{m}_2 + L_1 \vec{u}_{DF}, \notag \\
\vec{o}_2 &= \vec{n}_2 + \frac{\sqrt{3}}{2}L_1 \vec{u}_{DF} + \frac{L_1}{2} \vec{u}_{DF}^\perp, \notag \\
\vec{p}_2 &= \vec{n}_2 + \sqrt{3} L_1 \vec{u}_{DF}, \notag \\
\vec{q}_2 &= \vec{a}_2 + \frac{\sqrt{3}}{2} L_1 \vec{u}_{FB} - \frac{L_1}{2} \vec{u}_{FB}^\perp, \notag \\
\vec{r}_2 &= \vec{q}_2 + L_1 \vec{u}_{FB}, \notag \\
\vec{s}_2 &= \vec{r}_2 + L_1 \vec{u}_{FB}, \notag \\
\vec{t}_2 &= \vec{f}_2 + \frac{\sqrt{3}}{2} L_1 \vec{u}_{BC} - \frac{L_1}{2} \vec{u}_{BC}^\perp, \notag \\
\vec{u}_2 &= \vec{i}_2 + \frac{\sqrt{3}}{2} L_1 \vec{u}_{CD} - \frac{L_1}{2} \vec{u}_{CD}^\perp, \notag \\
\vec{v}_2 &= \vec{u}_2 + L_1 \vec{u}_{CD}, \notag \\
\vec{w}_2 &= \vec{v}_2 + L_1 \vec{u}_{CD}, \notag \\
\vec{x}_2 &= \vec{n}_2 + \frac{\sqrt{3}}{2} L_1 \vec{u}_{DF} - \frac{L_1}{2} \vec{u}_{DF}^\perp, \notag \\
\vec{y}_2 &= \vec{q}_2 + \frac{L_1}{2} \vec{u}_{FB} - \frac{\sqrt{3}}{2} L_1 \vec{u}_{FB}^\perp, \notag \\
\vec{z}_2 &= \vec{y}_2 + L_1 \vec{u}_{FB}, \notag \\
\vec{aa}_2 &= \vec{z}_2 + L_1 \vec{u}_{BC}, \notag \\
\vec{bb}_2 &= \vec{aa}_2 + L_1 \vec{u}_{CD}.
\label{eqn:appendix_square_deflation}
\end{align}

For yellow triangular tiles:
\begin{align}
\vec{a_3} &= \vec{F}, \notag \\
\vec{b_3} &= \vec{a_3} + L_1 \vec{u}_{FD}, \notag \\
\vec{c_3} &= \vec{b}_3 + \frac{\sqrt{3}}{2} L_1 \vec{u}_{FD} + \frac{L_1}{2} \vec{u}_{FD}^\perp, \notag \\
\vec{d_3} &= \vec{b_3} + \sqrt{3} L_1 \vec{u}_{FD}, \notag \\
\vec{e_3} &= \vec{D}, \notag \\
\vec{f_3} &= \vec{e_3} + L_1 \vec{u}_{DE}, \notag \\
\vec{g_3} &= \vec{f}_3 + \frac{\sqrt{3}}{2} L_1 \vec{u}_{DE} + \frac{L_1}{2} \vec{u}_{DE}^\perp, \notag \\
\vec{h_3} &= \vec{f_3} + \sqrt{3} L_1 \vec{u}_{DE}, \notag \\
\vec{i_3} &= \vec{E}, \notag \\
\vec{j_3} &= \vec{i_3} + L_1 \vec{u}_{EF}, \notag \\
\vec{k_3} &= \vec{j}_3 + \frac{\sqrt{3}}{2} L_1 \vec{u}_{EF} +\frac{L_1}{2} \vec{u}_{EF}^\perp, \notag\\
\vec{l_3} &= \vec{j_3} + \sqrt{3} L_1 \vec{u}_{EF}, \notag \\
\vec{m_3} &= \vec{k_3} - L_1 \vec{u}_{EF}^\perp, \notag \\
\vec{n_3} &= \vec{c_3} - L_1 \vec{u}_{FD}^\perp, \notag \\
\vec{o_3} &= \vec{g_3} - L_1 \vec{u}_{DE}^\perp
\label{eqn:appendix_triangle_y_deflation}
\end{align}

The iterative deflation process is implemented in Python based on the previously established mathematical relationships. Each tile appearing in the structure is stored as a nested tuple. The first element indicates the tile type—either \texttt{'square'}, \texttt{'triangle\_o'} (orange triangle), or \texttt{'triangle\_y'} (yellow triangle)—while the subsequent elements (three or four, depending on the shape) contain the nodal coordinates of the tile’s vertices. For example, the orange triangle in the base unit cell is stored as:

\begin{eqnarray}
    [((\text{'triangle\_o'}),(a_x,a_y),(b_x,b_y),(f_x,f_y))]
    \label{eqn:appendix_unit_cell_geom_format}
\end{eqnarray}

After applying the deflation rule, this triangle is replaced by a set of smaller tiles, represented as follows. Here, the subscripts $x$ and $y$ refer to the $\hat{\imath}$ and $\hat{\jmath}$ components of the corresponding vectors:

\begin{align}
    &[((\text{'triangle\_o'}),(d_{1,x},d_{1,y}),(c_{1,x},c_{1,y}),(o_{1,x},o_{1,y})), \notag \\
    &((\text{'triangle\_o'}),(a_{1,x},a_{1,y}),(l_{1,x},l_{1,y}),(m_{1,x},m_{1,y})), \notag \\
    &((\text{'triangle\_o'}),(m_{1,x},m_{1,y}),(l_{1,x},l_{1,y}),(k_{1,x},k_{1,y})), \notag \\
    &((\text{'triangle\_o'}),(q_{1,x},q_{1,y}),(i_{1,x},i_{1,y}),(h_{1,x},h_{1,y})), \notag \\
    &((\text{'triangle\_o'}),(g_{1,x},g_{1,y}),(q_{1,x},q_{1,y}),(h_{1,x},h_{1,y})), \notag \\
    &((\text{'triangle\_o'}),(d_{1,x},d_{1,y}),(o_{1,x},o_{1,y}),(e_{1,x},e_{1,y})), \notag \\
    &((\text{'square'}),(o_{1,x},o_{1,y}),(c_{1,x},c_{1,y}),(b_{1,x},b_{1,y}),(n_{1,x},n_{1,y})), \notag \\
    &((\text{'square'}),(n_{1,x},n_{1,y}),(b_{1,x},b_{1,y}),(a_{1,x},a_{1,y}),(m_{1,x},m_{1,y})), \notag \\
    &((\text{'square'}),(m_{1,x},m_{1,y}),(k_{1,x},k_{1,y}),(j_{1,x},j_{1,y}),(x_{1,x},x_{1,y})), \notag \\
    &((\text{'square'}),(f_{1,x},f_{1,y}),(p_{1,x},p_{1,y}),(q_{1,x},q_{1,y}),(g_{1,x},g_{1,y})), \notag \\
    &((\text{'square'}),(q_{1,x},q_{1,y}),(n_{1,x},n_{1,y}),(j_{1,x},j_{1,y}),(i_{1,x},i_{1,y})), \notag \\
    &((\text{'square'}),(x_{1,x},x_{1,y}),(j_{1,x},j_{1,y}),(i_{1,x},i_{1,y}),(q_{1,x},q_{1,y})), \notag \\
    &((\text{'triangle\_y'}),(n_{1,x},n_{1,y}),(m_{1,x},m_{1,y}),(x_{1,x},x_{1,y})), \notag \\
    &((\text{'triangle\_y'}),(o_{1,x},o_{1,y}),(n_{1,x},n_{1,y}),(p_{1,x},p_{1,y})), \notag \\
    &((\text{'triangle\_y'}),(p_{1,x},p_{1,y}),(n_{1,x},n_{1,y}),(x_{1,x},x_{1,y})), \notag \\
    &((\text{'triangle\_y'}),(p_{1,x},p_{1,y}),(x_{1,x},x_{1,y}),(q_{1,x},q_{1,y})) ]
    \label{eqn:appendix_sub_cell_geom_format}
\end{align}

By replacing each tile in the base unit cell (Eq.~\ref{eqn:appendix_unit_cell_geom_format}) with its corresponding set of substituted tiles (Eq.~\ref{eqn:appendix_sub_cell_geom_format}), we can construct the geometry of the deflated QC structure. This process is inherently iterative, enabling higher-order deflation levels by applying the same substitution rules repeatedly. The only limitation to this recursion is the available computational resources at the time. This data format for representing the geometry of QC structures is adapted from~\cite{gouldsbrough_computing_2021}.

\newpage
\section{\label{appendix:bandgap_identification} Stopband Identification with \textit{DBSCAN}}

\setcounter{figure}{0}
\setcounter{equation}{0}
\setcounter{table}{0}
\renewcommand{\thefigure}{B.\arabic{figure}}
\renewcommand{\theequation}{B.\arabic{equation}}
\renewcommand{\thetable}{B.\arabic{table}}

\subsection{Algorithm and Parameter Tuning}
\label{appendix:bandgap_identification_parameter}
\begin{figure}[h]
    \centering
    \includegraphics[width=\linewidth]{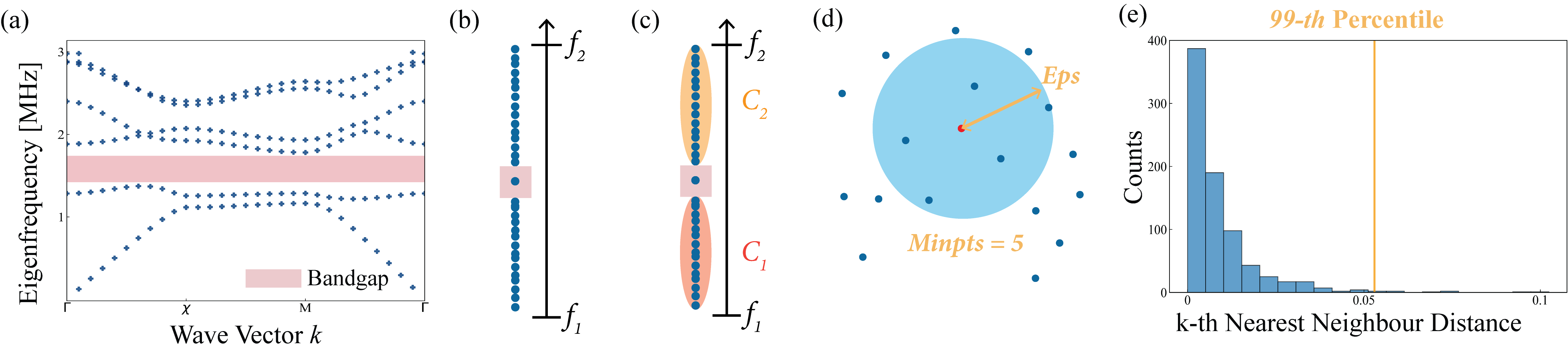}
    \caption{(a) Conventional bandstructure found for PnC structures, obtained via unit cell analysis with superimposed periodic boundary conditions. (b) Illustration of the 1D eigenfrequency array obtained for our QC-based designs, and possible definition of a stopband coming from data-driven perspectives. (c) Such 1D eigenfrequency array can be clustered based on the contrast in data density ($C_1$ and $C_2$), the distance in between neighbouring clusters indicates candidate stopband. (d) Conceptual definition of the key parameters \textit{MinPts} and \textit{Eps} used in our DBSCAN algorithm. (e) An example of the k-th nearest neighbour distance density distribution ($k=5$). 99th percentile of this distribution gives $\textit{Eps}\approx 0.05$.  }
    \label{fig:appendix_dbscan}
\end{figure}

In traditional periodic phononic crystal(PnC)-based resonator design, the typical approach involves analyzing a unit cell and deriving its bandstructure (Fig.~\ref{fig:appendix_dbscan}(a)) using Floquet periodic boundary conditions based on Bloch's theorem. From the resulting bandstructure, stopbands can be readily identified as frequency ranges where no Bloch modes exist. However, this approach is not directly applicable to our quasicrystal(QC)-based designs due to the lack of strict periodicity, which invalidates the assumptions underlying Bloch’s theorem and precludes the conventional unit-cell analysis.

Instead of using unit cell analysis, we performed a full eigenfrequency analysis on a symmetry-reduced slice of our design and obtained full spectrum of eigenfrequencies that serve as the foundation for our stopband identification process. Analogous to the unit-cell-based workflow, we interpret regions where no eigenfrequencies are present as candidate stopbands. Motivated by this density spectral perspective, we applied a one-dimensional density-based clustering algorithm to identify frequency intervals between high mode densities where stopbands are likely to appear in regions of low mode density (Fig.~\ref{fig:appendix_dbscan}(b)). The gaps between consecutive high-density clusters are then interpreted as potential stopbands (Fig.~\ref{fig:appendix_dbscan}(c)).

We have chosen to use the density-based spatial clustering of applications with noise (DBSCAN) algorithm~\cite{ester_density-based_1996}. The key advantage of DBSCAN is that it is non-parametric, requiring no prior assumptions about the underlying data distribution. Additionally, DBSCAN inherently identifies outliers by treating them as noise, this makes it a hard clustering algorithm where each data point is assigned to at most one cluster while allowing the presence of unclustered points~\cite{yin_rapid_2024}. This characteristic is particularly advantageous for stopband identification, as sparse eigenmodes within a stopband region can be safely excluded from the clustering result.

Although DBSCAN is a non-parametric algorithm, it still requires two input parameters: \textit{Eps} and \textit{MinPts}. A detailed explanation of the algorithm can be found in~\cite{ester_density-based_1996}. Here, we focus on the significance of these two parameters and the strategy used to define them in our context. Conceptually, DBSCAN iterates through each data point and counts the number of neighboring points within a specified distance \textit{Eps}. If this number exceeds the threshold defined by \textit{MinPts}, the point is considered part of a high-density region and is grouped accordingly. An illustration of this concept is shown in Fig.~\ref{fig:appendix_dbscan}(d). 

In our specific case, \textit{Eps} corresponds to the minimum tolerated distance between two consecutive high-density clusters in the frequency domain, or between a high-density cluster and isolated eigenmodes within a potential stopband. This is proportional to minimum stopband width of interest. Meanwhile, \textit{MinPts} defines the maximum number of eigenmodes we allow within a candidate stopband region before defining it as no longer a stopband.

Based on this physics-informed understanding, we define the DBSCAN parameter \textit{MinPts} as 5, since the number of eigenmodes found within a stopband is usually fewer than five. The definition of \textit{Eps}, however, requires a more nuanced approach. To adapt \textit{Eps} for each individual design, we employ a dynamic tuning strategy based on the density distribution of distances to the $k$-th nearest eigenmodes. For each eigenmode in the dataset, we compute the distance to its $k$-th nearest neighbor and construct the density distribution of these distances.

In such a distribution, the lower end reflects closely packed modes in high-density regions, while the higher end corresponds to larger distances between clusters or between clusters and isolated eigenmodes within stopband (see Fig.~\ref{fig:appendix_dbscan}(e)). When choosing $k \approx \textit{MinPts}$, the higher end of the distribution provides an estimate of the typical spacing associated with candidate stopbands in the design. Therefore, we define \textit{Eps} as the 99th percentile of this distribution. Given the approximate of number of eigenmodes found is usually between 600-800, 99th percentile returns the top 6-8 eigenmodes that are sparsely distributed. This is twice the number of eigenmodes usually found within a stopband region, making the approximation of \textit{Eps} smaller than the limiting value. 

This methodology helps us avoid assigning a \textit{MinPts} value that is too high, which could result in the merging of distinct clusters and, consequently, the loss of potential stopbands. On the other hand, underestimating \textit{MinPts} is less critical, as it tends to fragment clusters and yield more candidate stopbands. While this may increase computational time, it does not compromise the identification of key physics.

\subsection{Stopband Behavior of Design 2}

\begin{figure}[h]
    \centering
    \includegraphics[width=\linewidth]{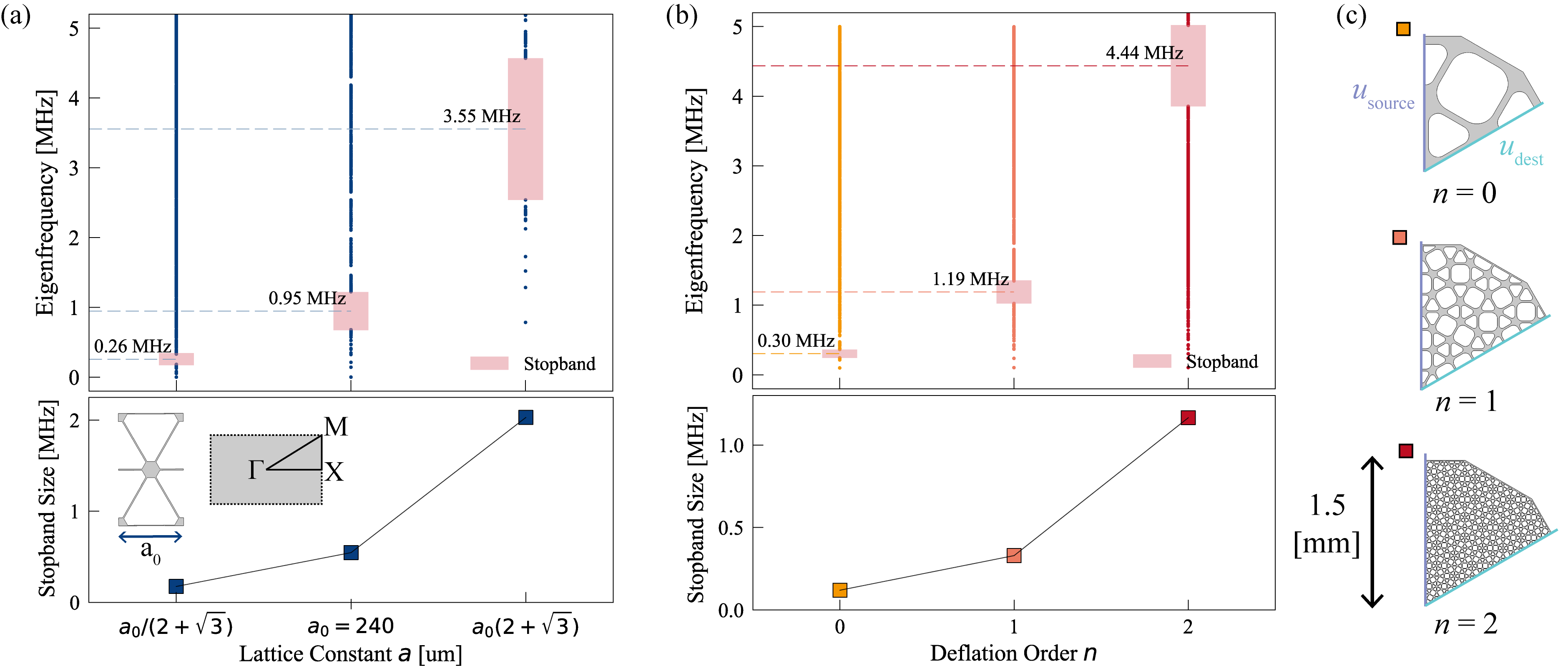}
    \caption{
\textbf{Comparison of stopband behavior between periodic phononic crystal and quasicrystal nanomechanical resonator designs for Design~2.}
(a) Stopband evolution in phononic crystal-based designs as a function of lattice constant $a_0$ ($r = 0.26a_0$). 
(b) Stopband evolution in quasicrystal-based designs as a function of deflation order $n$. 
(c) Representative quasicrystal geometries for $n=0,1,2$, illustrating the geometric refinement and the symmetry-reduced sector with cyclic boundary conditions.
}
    \label{fig:appendix_bandgap_HMC}
\end{figure}

In the main text, we have demonstrated how the stopband of our designs is influenced by the deflation order, $n$. Here, we analyze the stopband behavior of Design 2 (corresponding to Fig.~\ref{fig:bandgap_result}) and compare them with their periodic PnC counterparts.

For Design 2, where a stronger mass contrast is employed, we observe that both the stopband frequency and size increase as the characteristic length of the structure decreases, consistent with the trend seen in Design 1. However, the stopband size is generally larger compared to Design 1, indicating a potentially enhanced acoustic isolation capability. This observation aligns with the results reported in~\cite{reetz_analysis_2019}.

\subsection{Performance Validation}
\begin{figure}[h]
    \centering
    \includegraphics[width=\linewidth]{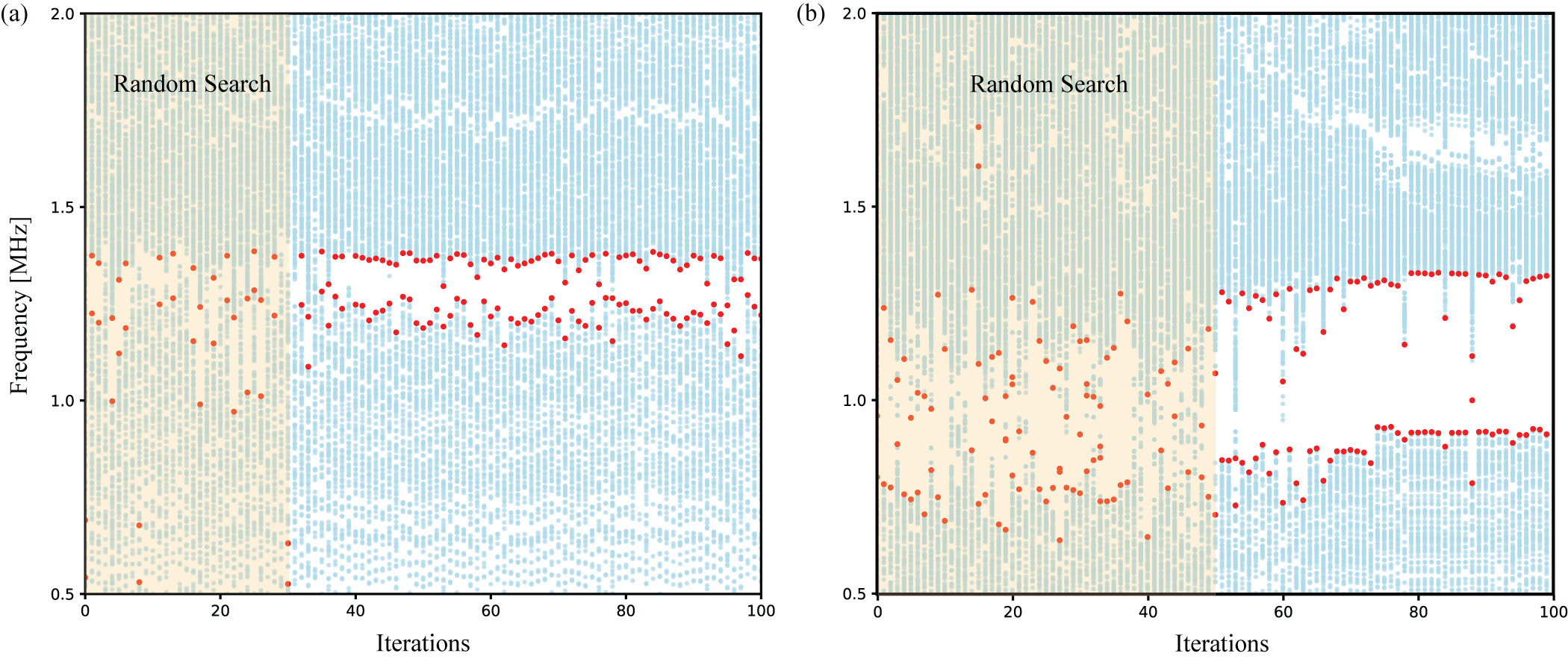}
    \caption{\textbf{Identified stopband boundaries within an optimization cycle of 100 design iterations for} (a) Design 1 (b) Design 2}
    \label{fig:appendix_dbscan_validation}
\end{figure}

To validate our proposed stopband identification algorithm using DBSCAN, we present the stopband identification results for Designs 1 and 2 across 100 design iterations in Fig.~\ref{fig:appendix_dbscan_validation}. Each column of blue dots represents the eigenfrequencies found for a specific design iteration, while the red dots denote the lower and upper boundaries of the identified stopbands.

Given the nature of Bayesian Optimization, the initial iterations consist of random exploration within the design space, where many poorly designed stopband behaviors can be observed. Beyond this random search region, the identified stopbands appear much cleaner, indicating that the implemented Bayesian Optimization inherently favors well-isolated stopband behaviors that correlate with improved $Q_m$ performance. Furthermore, Design~2 exhibits wider stopbands, suggesting better isolation, as discussed in~\cite{reetz_analysis_2019}.

\subsection{Frequency Response Validation}
\begin{figure}[h]
    \centering
    \includegraphics[width=\linewidth]{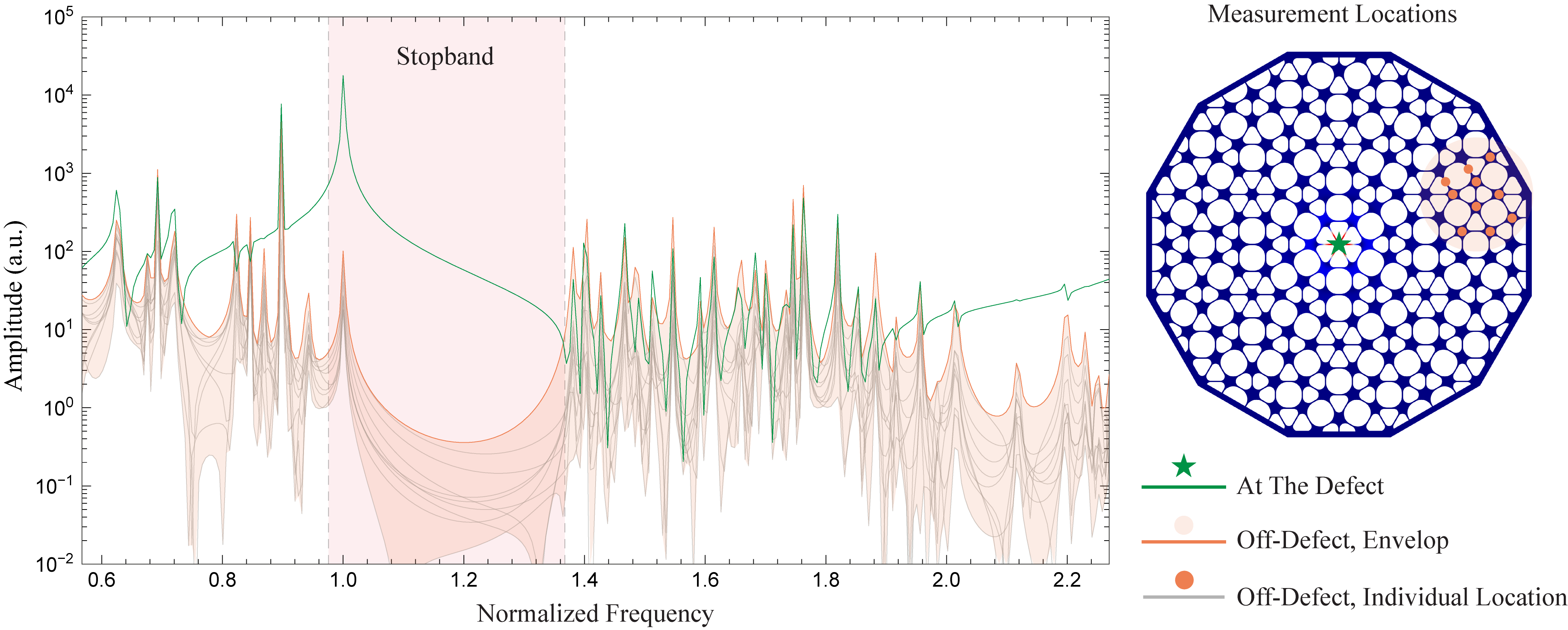}
    \caption{\textbf{Frequency-response analysis of Design~2 under harmonic excitation applied at the central defect.}}
    \label{fig:appendix_FR}
\end{figure}

In a finite quasicrystalline structure, a frequency interval with a low density of eigenmodes alone does not necessarily establish a phononic stopband. As described in Section~\ref{sec:2.A}, candidate stopband intervals were identified from the eigenfrequency spectrum and further assessed through mode localization. To provide more direct dynamical verification, we additionally performed a frequency-response analysis for the representative Design~2.
A harmonic excitation was applied at the central defect, and the responses were evaluated at the defect location and at ten additional sampling points extending progressively toward the clamped boundary. The excitation frequency was swept from $0.5$~MHz to $2$~MHz using 300 frequency points and normalized by the defect-mode resonance frequency, $f_0$. The results are presented in Fig.~\ref{fig:appendix_FR}.
Within the identified stopband, the off-defect responses are strongly suppressed, while a pronounced resonant response remains localized at the central defect near $f/f_0=1$. Outside the stopband, the off-defect responses exhibit multiple resonant peaks and are substantially less suppressed. These results provide frequency-domain evidence supporting wave isolation and defect-mode localization within the identified stopband.

\section{\label{appendix:design_parameters} Design Parameters}

\setcounter{figure}{0}
\setcounter{equation}{0}
\setcounter{table}{0}
\renewcommand{\thefigure}{C.\arabic{figure}}
\renewcommand{\theequation}{C.\arabic{equation}}
\renewcommand{\thetable}{C.\arabic{table}}

To realize the proposed nanomechanical resonator designs, we propose to use high-stress $\text{Si}_3\text{N}_4$ membranes with a thickness ($t$) of 35~nm and diameter of 3~mm. This is primarily attributed to the high initial pre-stress of approximately $1.27$~GPa in $\text{Si}_3\text{N}_4$, introduced by its deposition process. The pre-stress increases the stored elastic energy in the structure, which can lead to improved $Q_m$ performance. The material properties of $\text{Si}_3\text{N}_4$ are taken as $E = 270$~GPa, $\nu = 0.27$, $\rho = 3200$~kg/m$^3$.

Starting with the uniform string-based design, we used three design variables $w_y, w_o, w_s$ to control the width of the strings generated at the three distinct tiles, together with an additional design variable, $r_1$, controlling the radius of the fillet at the central defect pad, giving us 4~DoFs in the design space. In Design 1, we proposed $r_y, r_o, r_s$ to control the radius of the holes generated at distinct tiles. With slightly different central defect design, $r_1$ and $r_2$ are used to control the radius of the hole and the size of the central pad, respectively, resulting in a total of 5 design parameters. Design 2 has 7 independent design parameters. The radius of the fillets and width of strings at distinct tiles are controlled via $r_y, r_o, r_s$ and $w_y, w_o, w_s$, respectively. 

Furthermore, the minimum features in all designs are limited to the micrometer scale to enclose possibility for large-scale fabrication with state-of-the-art photolithography techniques~\cite{shin_spiderweb_2022}. It is also important to note that the defect designs proposed in this work are arbitrary and thorough defect engineering can have a profound impact on the performance of nanomechanical resonators~\cite{tsaturyan_ultracoherent_2017,halg_membrane-based_2021,engelsen_ultrahigh-quality-factor_2024}. In the absence of a more systematic design rationale, the defects in all our structures were introduced by either removing or adding material to form a central pad.

\section{\label{appendix:simulation} Finite-Element Simulation}

\setcounter{figure}{0}
\setcounter{equation}{0}
\setcounter{table}{0}
\renewcommand{\thefigure}{D.\arabic{figure}}
\renewcommand{\theequation}{D.\arabic{equation}}
\renewcommand{\thetable}{D.\arabic{table}}

\subsection{Simulation Setup}
\begin{figure}[t]
    \centering
    \includegraphics[width=\linewidth]{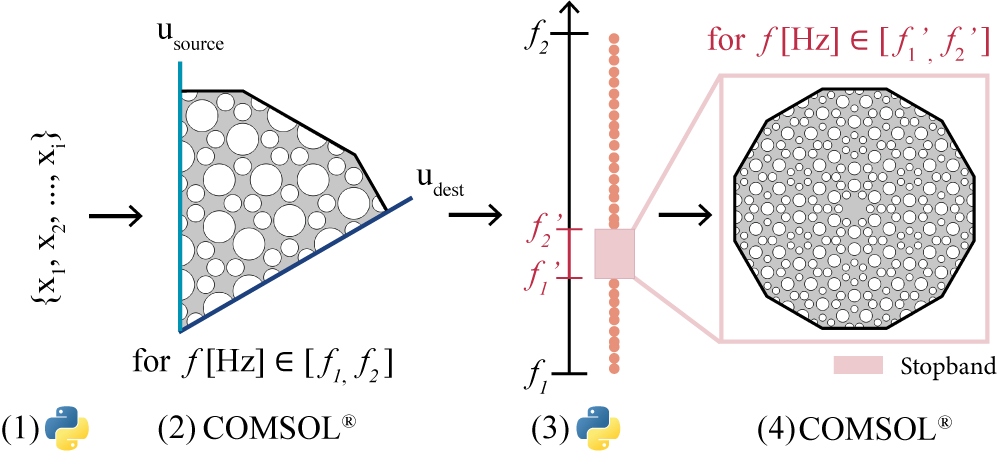}
    \caption{\textbf{Simulation pipeline for the data-driven design of a quasicrystal-based nanomechanical resonator. }
    A symmetry-reduced eigenfrequency analysis is used to identify stopbands, after which full-structure simulations are performed within the selected frequency windows to evaluate localized modes and their mechanical quality factors $Q_m$.}
    \label{fig:appendix_simulation_pipeline}
\end{figure}

Our device is simulated using a prestressed eigenfrequency study, employing plate physics within COMSOL. As outlined in Fig.~\ref{fig:appendix_simulation_pipeline}, the simulation pipeline consists of two independent simulations: one focused on identifying the stopband and the other on estimating the $Q_m$ performance.

The first simulation is carried out on a reduced model exploiting the cyclic symmetry of the 12-fold quasicrystal (QC) structure. This corresponds to a slice with an angular span of $\pi/3$. With the application of Floquet cyclic-symmetric boundary conditions (Eq.~\ref{eqn:cyclic_periodic_bc}), we use the ARPACK eigenfrequency solver to extract eigenmodes in the frequency range of 0.5–3~MHz. Lower frequency modes are excluded, as our focus lies on higher-order modes. Due to the symmetry condition, we sweep over the azimuthal mode number from 0 to 3 (integers only), setting the maximum number of eigenmodes per sweep to 500, approximate number of eigenmodes to 200. The eigenmodes obtained from this step are used to identify stopbands, as detailed in Supporting Information~\ref{appendix:bandgap_identification}.

The second simulation is performed on the full device geometry but restricted to the frequency range between the lower ($f_1'$) and upper ($f_2'$) stopband boundaries identified in the first step. To enhance computational efficiency, we employ an around shift eigenfrequency solver rather than a full search region solver. We target an approximate number of 20 eigenmodes, centered around a shift frequency of $(f_1'+f_2')/2$. This ensures that, a maximum of 20 defect modes within the stopband, will be efficiently captured.

\subsection{Mesh Configuration and Study}
Another key setting in our simulation is the mesh configuration. Since bending loss is related to the second-order derivatives of the out-of-plane displacement field (Eq.~\ref{eqn:Q_calculation}), the computed $Q_m$ values in finite-element simulations are known to be sensitive to mesh size—especially in regions exhibiting significant bending. Therefore, selecting an appropriate mesh size and distribution is critical for achieving both accurate results and efficient design optimization.

To evaluate the impact of mesh resolution on the $Q_m$ performance, we conducted mesh convergence studies independently for two representative designs (Fig.~\ref{fig:design_formulation_method}). For each design, we examined three typical mode shapes: (1) a well-isolated defect mode, (2) an isolated defect mode with minor leakage, and (3) a global mode. This provides a comprehensive view of the mesh sensitivity across different dynamical behaviors. It is worth noticing that the mesh configuration discussed in the following section is only used for the full-mode simulation.

\subsubsection{Design 1}
\begin{figure}[h]
    \centering
    \includegraphics[width=0.95\linewidth]{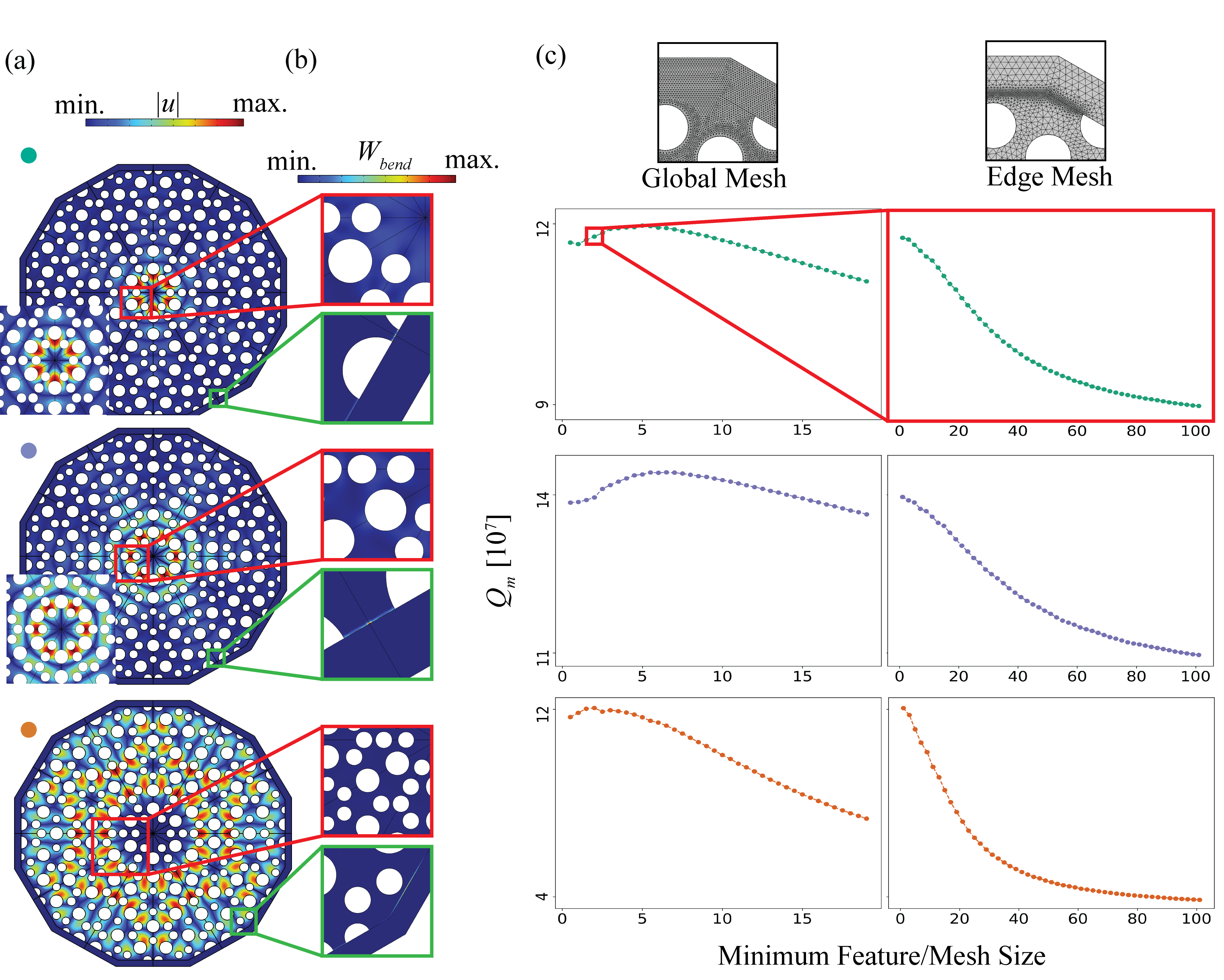}
    \caption{\textbf{Mesh convergence analysis for Design 1 over different characteristic eigenmodes and mesh configurations.}
    (a) Three common characteristic eigenmodes seen, color map indicates the magnitude of out-of-plane displacement. (From top to bottom) A well isolated defect mode, a less isolated defect mode, and a global mode. (b) Distribution of the energy dissipated per cycle due to bending loss, \(W_{\rm bend}\), in each individual eigenmode. Strong localization is observed close to the clamped boundary. (c) Effects of reducing mesh size on $Q_m$ performance for global and edge mesh configurations. The mesh size is defined relative to the minimum structural feature found in the design.}
    \label{fig:appendix_massive_LMC_mesh}
\end{figure}

From the results shown in Fig.~\ref{fig:appendix_massive_LMC_mesh}, we observe minor bending loss distributed across the membrane in regions exhibiting significant curvature. To study the effect of mesh on $Q_m$ performance, we first started by gradually reducing the global mesh size where we could observe a substantial decrease in $Q_m$ performance. However, due to computational limitations, we were only able to reduce the global mesh size to approximately 20 times smaller than the smallest structural feature. To obtain more reliable results, a more efficient mesh configuration is needed.

Closer inspection of the mode shapes reveals that none of the three modes are perfectly localized, with some vibration energy propagating toward the boundaries. By refining the mesh specifically at the clamped boundary, we noticed significant bending loss localization at the clamped boundary, as illustrated in Fig.~\ref{fig:appendix_massive_LMC_mesh}(b). When sweeping the edge mesh size, a clear drop in $Q_m$ is observed for all three modes, especially for the global mode, clearly capturing the same trend seen in the global mesh convergence study. Here, we see the convergence trends emerging at a mesh size approximately 80 times smaller than the smallest structural feature. During this edge mesh study, all other regions of the geometry were meshed with a size equivalent to half of the smallest feature.

Given that the minimum feature size in our design space can be as small as $6\,\mu\text{m}$, using such fine meshes throughout would be computationally prohibitive. Therefore, for simulations within the optimization loop, we employ a finer edge mesh domain with size of $0.3\,\mu\text{m}$—approximately 20 times smaller than the minimum feature in the most geometrically demanding designs—while using a global mesh size at least twice smaller than the minimum feature for the rest of the geometry. This configuration is sufficient to correctly distinguish between lossy global modes and well-confined defect modes, which might otherwise be overestimated as high-$Q_m$ modes if coarse meshes are used. For final evaluations, the $Q_m$ performance of the optimized design is validated using a converged edge mesh approximately 80 times smaller than the minimum feature size.

\subsubsection{Design 2}
In this design, we observed significant bending at the edges of the pads located between interconnected strings. To assess the impact of mesh size on the $Q_m$ performance, we first conducted a convergence study by refining the global mesh. Due to computational limitations, we were only able to reduce the global mesh size down to approximately 1/13 of the minimum structural feature. Nevertheless, convergence trend in $Q_m$ was observed for all three modes when the mesh size reached below 1/6 of the minimum feature size.

To further reduce computational cost, we applied a locally refined mesh along the filleted edges that form the pads, while maintaining a coarser global mesh at approximately half of the minimum structural feature size. For well-isolated modes (e.g., Mode 1), this configuration was already sufficient to capture most of the bending losses, with convergence observed for a fillet mesh size around six to eight times smaller than the minimum structural feature.

However, for modes where vibrations extend toward the clamped boundary (Modes 2 and 3), this was insufficient to resolve the sharp bending losses localized at the fixed supports. To address this, we introduced a third refinement region with a dedicated fine mesh along the clamped edge, while keeping the fillet mesh fixed at the previously optimized size. This edge refinement revealed substantial reductions in $Q_m$ for Modes 2 and 3, with Mode 3 showing the most significant drop. Convergence was observed for an edge mesh size around ten times smaller than the minimum structural feature. Based on these observations, we adopted a three-domain meshing strategy for simulations within the optimization loop: a global (rough) mesh at half, a fillet mesh at six times, and an edge mesh at eight times of the minimum structural feature size. And we will be more confident with the $Q_m$ result if an isolated mode is returned, the effects of mesh are much less significant.

\begin{figure}[h]
    \centering
    \includegraphics[width=\linewidth]{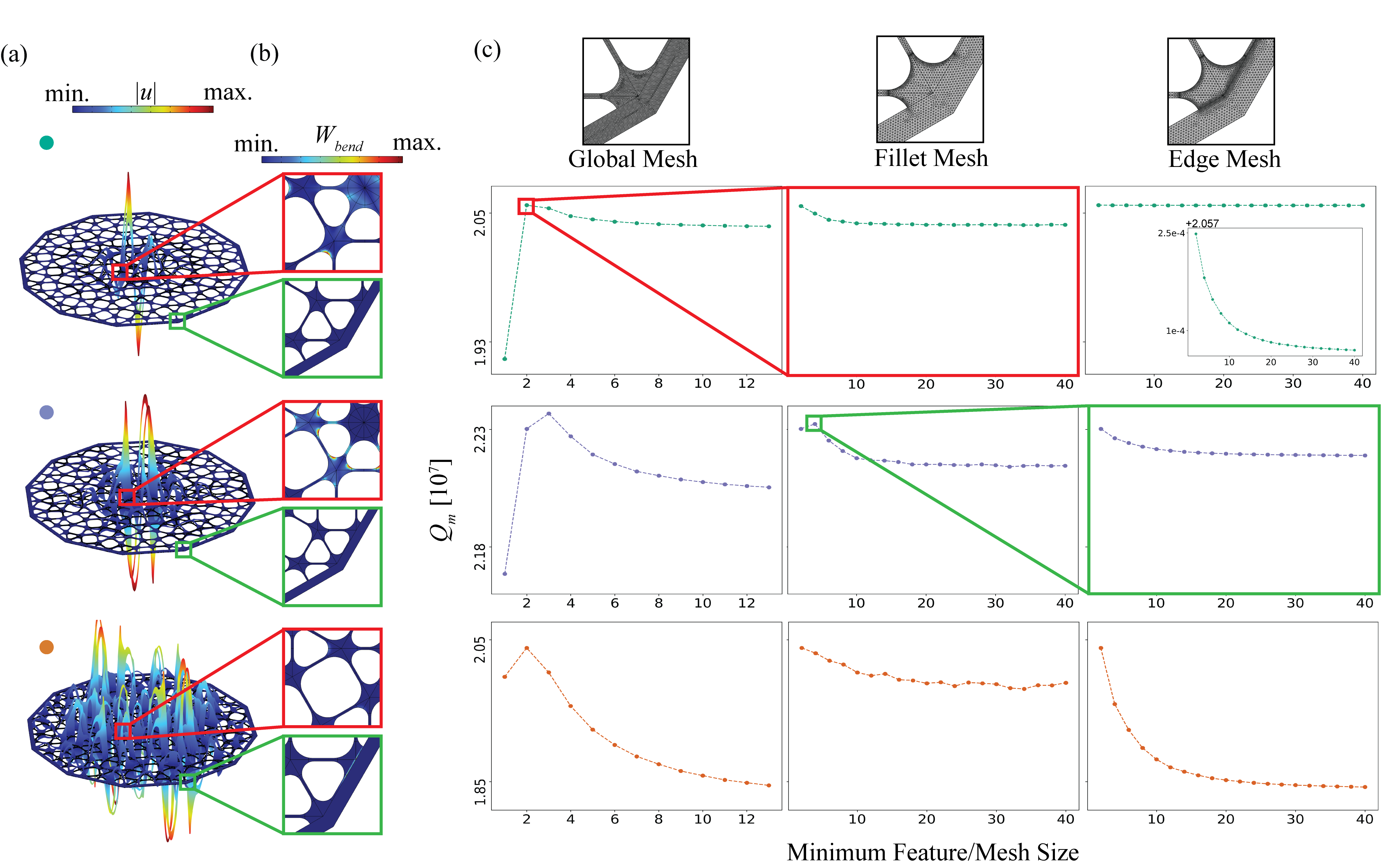}
    \caption{\textbf{Mesh convergence analysis for Design 2 over different characteristic eigenmodes and mesh configurations.} (a) Three common characteristic eigenmodes seen, color map indicates the magnitude of out-of-plane displacement. (From top to bottom) A well isolated defect mode, a less isolated defect mode, and a global mode. (b) Distribution of the energy dissipated per cycle due to bending loss, \(W_{\rm bend}\), in each individual eigenmode. Strong localization is observed close to the fillet edges and clamped boundary for the global mode. (c) Effects of reducing mesh size on $Q_m$ performance for global, fillet and edge mesh configurations. The mesh size is defined relative to the minimum structural feature found in the design.}
    \label{fig:appendix_massive_HMC_mesh}
\end{figure}

\section{\label{appendix:bayesian_optimization} Bayesian Optimization Formulation}

\setcounter{figure}{0}
\setcounter{equation}{0}
\setcounter{table}{0}
\renewcommand{\thefigure}{E.\arabic{figure}}
\renewcommand{\theequation}{E.\arabic{equation}}
\renewcommand{\thetable}{E.\arabic{table}}

Bayesian optimization (BO) is an online machine learning algorithm that learns cumulatively from all previous steps (known as the prior model, $D_i$) and makes probabilistic predictions on the objective function (known as the posterior model, $f_i(x,D_i)$) based on Gaussian Processes (GP). The selection of the next step is determined by the posterior model and an acquisition function ($\alpha_i(x,D_i)$). BO is particularly effective in dealing with black-box optimization and expensive simulations, and it is well-known for reaching global optima with a minimal number of steps \cite{bajaj_black-box_2021,shahriari_taking_2016}. Additionally, BO has various variants that can address multi-fidelity and multi-objective optimization problems, providing versatile solutions to our problem.

\begin{figure}[h]
    \centering
    \includegraphics[width=\linewidth]{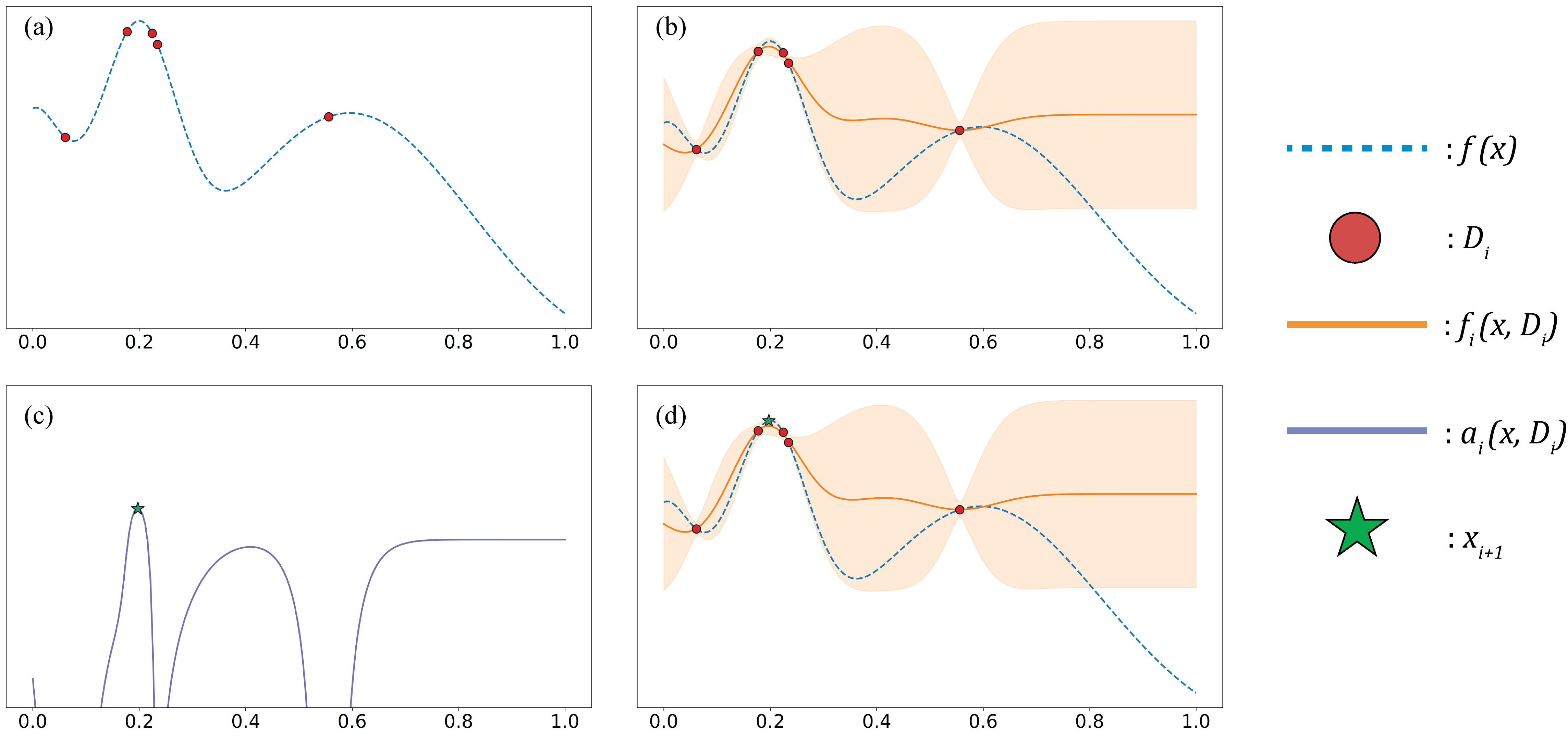}
    \caption{(a) The actual model for optimization. (b) Posterior model constructed after a few initial samplings. (c) An exemplar acquisition function (Expected Improvements) used. (d) The next best sample point obtained by maximizing the acquisition function in (c).}
    \label{fig:appendix_bayesian_opt}
\end{figure}

To further clarify the inner workings of BO, Fig.~\ref{fig:appendix_bayesian_opt} illustrates the key steps in a typical BO iteration. The process begins with the construction of an initial dataset (Fig.~\ref{fig:appendix_bayesian_opt}(a)), which serves as the prior model. This is followed by fitting a GP model to generate the posterior distribution (Fig.~\ref{fig:appendix_bayesian_opt}(b)). Unlike conventional regression methods, GPs provide not only a mean prediction but also an uncertainty estimate in the form of a full probability distribution. Based on this posterior, an acquisition function is evaluated (Fig.~\ref{fig:appendix_bayesian_opt}(c)).

In our implementation, we employ the logarithmic expected improvement (LogEI) as the acquisition function. LogEI is an enhancement of the conventional expected improvement (EI), addressing the vanishing gradient issue that often arises in traditional EI formulations~\cite{ament_unexpected_2025}. By maximizing this acquisition function, the algorithm selects the most promising candidate for evaluation (Fig.~\ref{fig:appendix_bayesian_opt}(c) and (d)).

The full procedure is summarized mathematically in Algorithm~\ref{alg:bayesian_opt}.

\begin{algorithm}
\caption{General Bayesian Optimization}
\label{alg:bayesian_opt}
\begin{algorithmic}[1]
    \State \textbf{Input:} Acquisition function $\alpha(x,D)$
    \State \textbf{Initialize:} $D_i = \{(x_1, y_1), \dots, (x_i, y_i)\}$ \Comment{Initial dataset}
    
    \For{$i = 1,2,\dots$}
        \State Fit $f_i(x, D_i)$ with Gaussian Process
        \State Compute acquisition function $\alpha_i(x, D_i)$
        \State Select next sample point $x_{i+1} = \arg\max_{x} \alpha_i(x, D_i)$
        \State Evaluate objective function at $x_{i+1}$
        \State Update: $D_{i+1} = D_i \cup \{(x_{i+1}, y_{i+1})\}$
    \EndFor
    
\end{algorithmic}
\end{algorithm}

\section{\label{appendix:other_studies} Other Interesting Properties of Quasicrystal-based Resonator Design}

\setcounter{figure}{0}
\setcounter{equation}{0}
\setcounter{table}{0}
\renewcommand{\thefigure}{F.\arabic{figure}}
\renewcommand{\theequation}{F.\arabic{equation}}
\renewcommand{\thetable}{F.\arabic{table}}

\subsection{In-plane Strain Engineering}
\label{appendix:other_studies_inplane_se}
\begin{figure}[h]
\centering
\includegraphics[width=\linewidth]{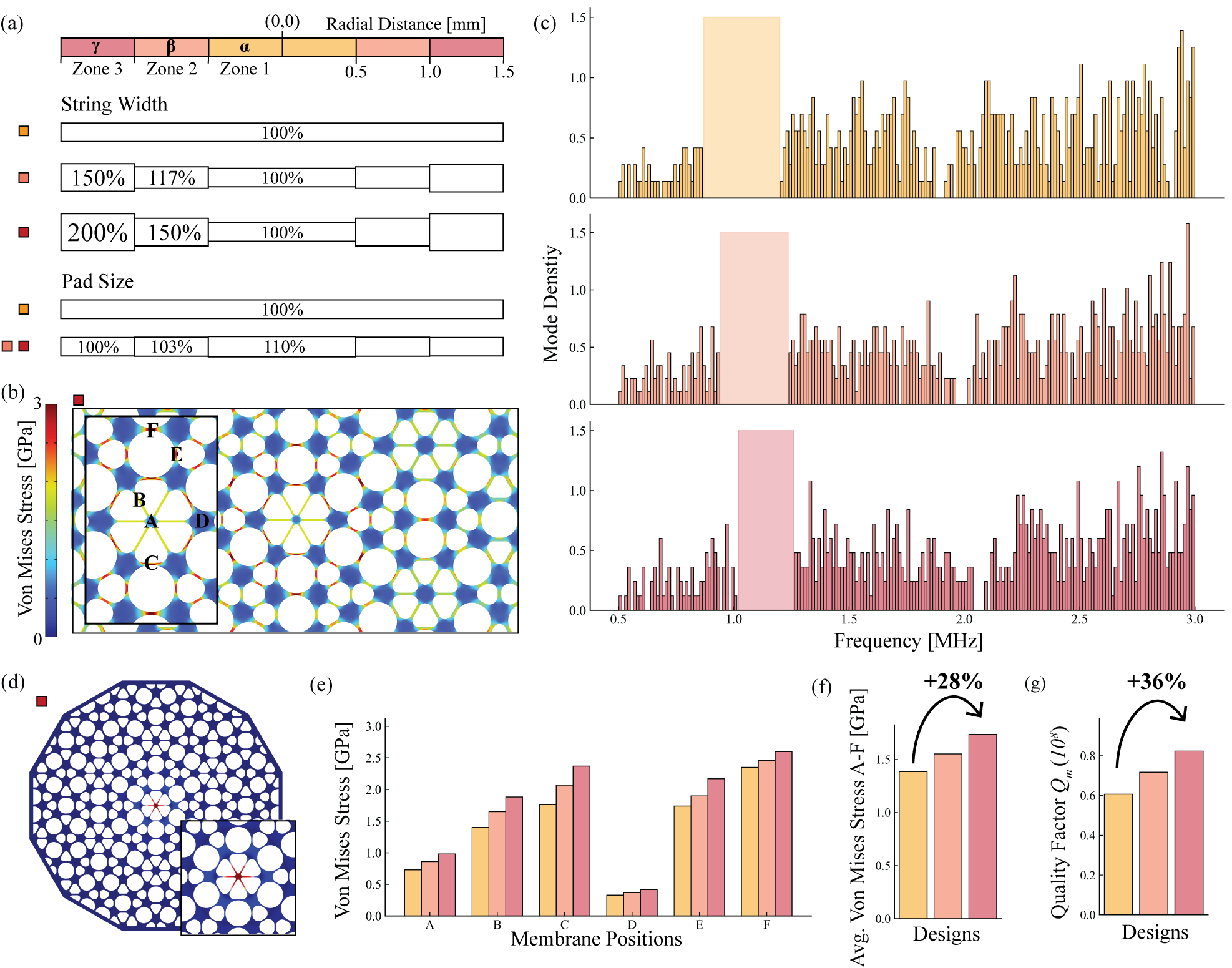}
\caption{\textbf{Demonstration of in-plane strain engineering techniques with optimized Design 2.} (a) Distribution of radial zones in enabling radial direction variation of design features. Shows the radial direction design feature profile for three separate designs, from no strain engineering to most extreme strain engineering accomplished. (b) Von Mises stress field near the defect for design with most strain engineering applied. (c) Frequency spectrum (simulated) for three designs, stopband regions are highlighted. (d) Soft-clamped mode shape of the most strain engineered design. (e) Averaged von Mises stress in regions A-F near the defect for three designs, labels in (b). (f) Averaged von Mises stress across all regions A-F for three designs. (g) Quality factor ($Q_m$) for three designs.}
\label{fig:in-plane_SE}
\end{figure}

Another key strength of quasicrystal(QC)-based resonator design is its structural tendency towards the center, ideal for in-plane strain engineering. Given the optimized designs presented in Fig.~\ref{fig:result_mode_shape}, we can validate the compatibility of QC-based resonator designs with in-plane strain engineering. Based on the best sensitivity performance achieved, we chose to implement in-plane strain engineering using the optimized Design 2.

To achieve this, we introduced three additional degrees of freedom in design along the radial direction. By dividing the membrane into three distinct radial zones, we can independently tune the design features based on their distance from the center. Three distinct radial profiles are shown in Fig.~\ref{fig:in-plane_SE}, corresponding to three different designs. The uniform profile represents the optimized design shown in Fig.~\ref{fig:result_mode_shape}, where no modifications are made to the design parameters along the radial direction. In the other two profiles, the string widths gradually decrease toward the center, while the pad size increases. Mathematically, considering the optimized string width ($w_{i,\mathrm{opt}}$) and pad size ($r_{i,\mathrm{opt}}$), where $i = \{y,o,s\}$ corresponds to the three different tilings, the design parameters after strain engineering ($w^{j}_{i,\mathrm{SE}}$ and $r^{j}_{i,\mathrm{SE}}$) can be defined as:

\begin{eqnarray}
    & \text{Zone 1} \rightarrow w^{1}_{i,\mathrm{SE}} = \alpha w_{i,\mathrm{opt}}, \hspace{10pt} r^{1}_{i,\mathrm{SE}} = \alpha r_{i,\mathrm{opt}} \\
    & \text{Zone 2} \rightarrow w^{2}_{i,\mathrm{SE}} = \beta w_{i,\mathrm{opt}}, \hspace{10pt} r^{2}_{i,\mathrm{SE}} = \beta r_{i,\mathrm{opt}} \\
    & \text{Zone 3} \rightarrow w^{3}_{i,\mathrm{SE}} = \gamma w_{i,\mathrm{opt}}, \hspace{10pt} r^{3}_{i,\mathrm{SE}} = \gamma r_{i,\mathrm{opt}}
    \label{eqn:se_design_parameters}
\end{eqnarray}

where $\alpha$, $\beta$, and $\gamma$ are the scaling factors corresponding to the radial direction degrees of freedom. With such strain engineering, we can increase the von Mises stress near the defect region up to 3~GPa, as shown by the localized stress region in Fig.~\ref{fig:in-plane_SE}(b). 

However, such a brute attempt at strain engineering can also lead to a mismatch in stopband frequency across different radial zones. This is because the stopband frequency is stress-dependent, regions with higher stress tend to exhibit higher stopband frequencies. This mismatch is evident in the frequency density spectrum (Fig.~\ref{fig:in-plane_SE}(c)), where the lower-frequency region of the stopband disappears under more extreme strain engineering. In conventional strain-engineered phononic crystals (PnCs), such an issue can be mitigated by reducing the lattice constant ($a_0$) in the low-stress regions to match the stopband frequency. In this early-stage validation, we disregarded the matching process and assumed that the stopband persists, which is supported by the observation of the local resonance defect modes (Fig.~\ref{fig:in-plane_SE}(d)). However, this mismatch fundamentally limits the extent of strain engineering that can be applied in our designs.

By computing the stress in multiple regions around the defect, we approximate an average increase of 28\% in von Mises stress near the defect. This mapped to a nearly 36\% increase in $Q_m$, yielding $Q_m = 0.82\times10^8$ at $f_m = 1$~MHz. While this represents a 54\% increase in the resonator's coherence, the effective mass remains as low as $0.97$~ng, resulting in a slightly improved thermal-force-noise-limited sensitivity of $\sqrt{S_{FF}} = 24.8$~aN/$\sqrt{\text{Hz}}$.

\newpage
\subsection{Effect of Device Size on In-plane Strain Engineering}
\label{appendix:other_studies_device_size}

Another interesting property of QC structures is their ability to grow infinitely in size by following the deflation rules defined in Fig.~\ref{fig:design_formulation_method}(a). With an increase in aspect ratio, we would naturally expect an improvement in $Q_m$ performance, as observed in extreme-aspect-ratio PnC-based designs~\cite{cupertino_centimeter-scale_2024}. But, beyond the benefits of a higher aspect ratio, we hypothesize that constructing a larger QC-based nanomechanical resonator could further amplify the effects of in-plane strain engineering, thereby increasing the stress level near the defect region.

\begin{figure}[h]
    \centering
    \includegraphics[width=0.625\linewidth]{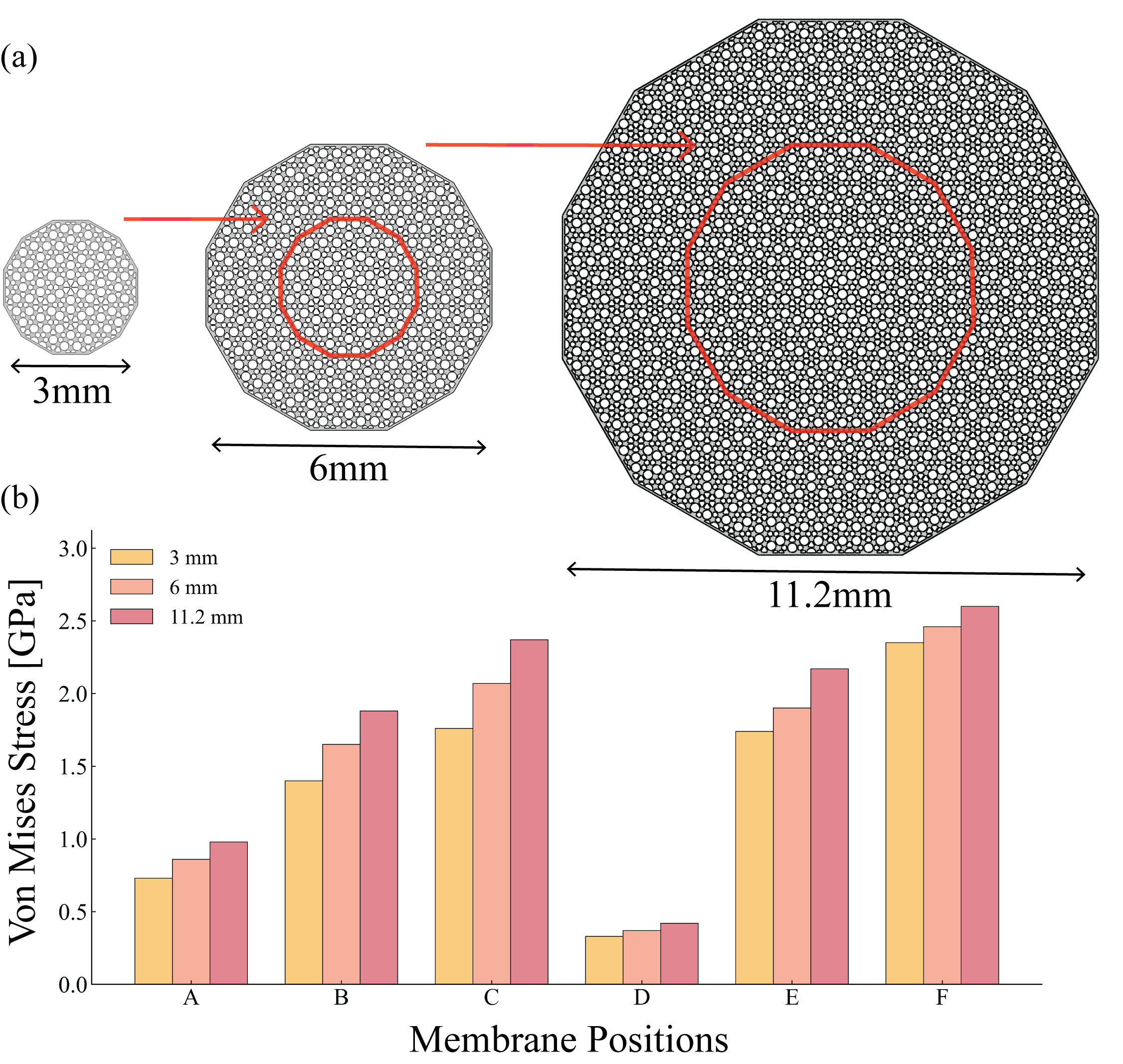}
    \caption{(a) Construction of our optimized Design 2 with the most extreme in-plane strain engineering implemented for different sizes: 3mm, 6mm, 11.2mm (from left to right). (b) Averaged von Mises stress in regions A-F (defined in Fig.~\ref{fig:in-plane_SE}b) for three designs with different sizes.}
    \label{fig:add_layer}
\end{figure}

To prove our hypothesis, we took the optimized Design~2 with the most extreme in-plane strain engineering implemented in Fig.~\ref{fig:in-plane_SE}, and added additional layers surrounding the original design to increase its size from 3~mm in diameter to 6~mm and 11.2~mm, consecutively (Fig.~\ref{fig:add_layer}). The added layers share identical design parameters with zone~3 of the strain-engineered design (Fig.~\ref{fig:in-plane_SE}(a)). Without any further design or optimization, we already observe a \(\sim 7\%\) increase in the redistributed von Mises stress field in regions A–F when doubling the resonator size to 6~mm. An additional 2\% increase in stress levels is seen when further increasing the resonator size. Although the gradient of stress enhancement decreases, we demonstrated the capability of QC-based resonators to further enhance strain engineering simply by adding additional layers of the crystal structure and increasing the resonator size. This also indicates the potential benefits that QC-based resonators could offer under extreme aspect ratio designs, which can be less feasible for conventional periodic designs.

\subsection{Robustness to Fabrication Variability}
\label{appendix:other_studies_robustness}
When nanomechanical resonators are used in optomechanical systems, the matching between the mechanical frequency and the optical resonance is critical to the overall system performance~\cite{aspelmeyer_cavity_2014}. In the following section, we present a preliminary study on the robustness of nanomechanical resonator designs to fabrication variability. Specifically, we focus on errors or variations in the geometric features of the fabricated device relative to the original design. In our case, we focus on the geometric features of Design~2, namely the beam width and fillet radius. To quantify this, we define the fractional drift in these geometric parameters as

\begin{eqnarray}
    & \delta_{w_i} = \frac{\varepsilon_{w_i}}{w_i}, \\
    & \delta_{r_i} = \frac{\varepsilon_{r_i}}{r_i},
    \label{eqn:appendix_fractional_drift}
\end{eqnarray}

where $w_i$ and $r_i$ correspond to the optimized design parameters for $i = \{y, o, s\}$, and $\delta$ and $\varepsilon$ represent the fractional and absolute drifts, respectively.

\begin{figure}[h]
    \centering
    \includegraphics[width=\linewidth]{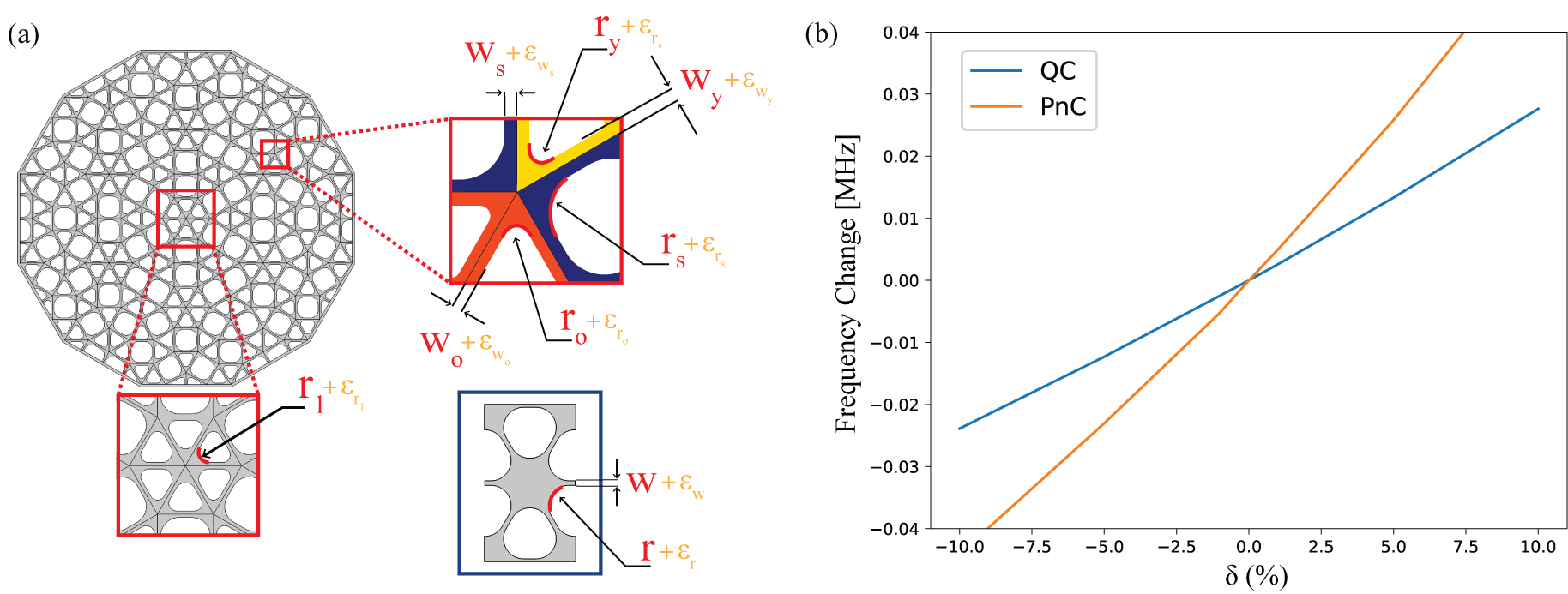}
    \caption{(a) Design parameters after introducing absolute drifts caused by fabrication variability. Note that the defect region in the PnC-based design also employs a fillet-based geometry, and the corresponding drift is applied accordingly (not shown in the figure). (b) Resonance frequency variation in both QC- and PnC-based nanomechanical resonators with respect to the fractional drift in the design parameters.}
    \label{fig:appendix_robustness}
\end{figure}

A positive drift could occur when excessive etching removes too much material, resulting in narrower beams and smaller fillet radii. Conversely, a negative drift occurs when insufficient etching leaves excess material, producing wider beams and larger fillet radii. By assigning different percentage drifts, we study their effects on the resonance frequency of the high-$Q_m$ mode in our optimized Design~2, corresponding to the mode shape shown in Fig.~\ref{fig:result_mode_shape}. Simulating the drifted geometries allows us to predict the frequency deviation in response to fabrication errors, as shown in Fig.~\ref{fig:appendix_robustness}. 

For comparison, the same fractional drifts were applied to a comparable PnC-based nanomechanical resonator design. It is clearly observed that our QC-based design is more robust against fabrication variability that alters geometric dimensions. This study serves as a preliminary investigation highlighting potential advantages of QC-based nanomechanical resonator designs, with further experimental validation needed in future work.

\section{\label{appendix:alt_qc_structure} Details on Alternative Quasicrystal Structures}

\setcounter{figure}{0}
\setcounter{equation}{0}
\setcounter{table}{0}
\renewcommand{\thefigure}{G.\arabic{figure}}
\renewcommand{\theequation}{G.\arabic{equation}}
\renewcommand{\thetable}{G.\arabic{table}}

As mentioned in Fig.~\ref{fig:introduction}, numerous types of quasicrystals (QCs) exist, and in this work we additionally studied performance of resonator design based on two alternative QC structures. The deflation rules for the alternative QC structures is highlighted in Fig.~\ref{fig:appendix_alt_qc_structures}(a). As with the 12-fold design, we employ high-stress $\mathrm{Si_3N_4}$ membranes with a thickness of 35~nm and a diameter of 3~mm, and the pre-stress as well as the material properties of $\mathrm{Si_3N_4}$ are kept fully consistent with those used in the original 12-fold design. Using this approach, we observe mechanical stopbands in these alternative QC geometries and, after optimization, confirm the emergence of soft-clamped modes that are strongly localized near the defect region. The key performances of these design are provided in Table~\ref{tab:appendix_alt_qc_structure}.

\begin{table}[h]
\caption{\label{tab:appendix_alt_qc_structure}Simulated performance of non-12 fold Designs assuming room temperature ($T = 300\,\mathrm{K}$). Designs are defined in Fig.~\ref{fig:appendix_alt_qc_structures}(b)}
\begin{tabular}{l|c|c|c|c}
\textrm{Designs} & \textrm{3} & \textrm{4} & \textrm{5} & \textrm{6}  \\
\hline
Frequency $f$ (MHz) & 1.52 & 1.50 & 0.83 & 0.92 \\
Effective mass $m_\mathrm{eff}$ (ng) & 2.8 & 1.11 & 21.7 & 3.1 \\
Quality factor $Q_m$ ($\times 10^6$) & 63.1 & 41.1 & 140 & 99.7 \\
$Qf$ product (THz) & 96 & 61.7 & 117 & 91.3 \\
\end{tabular}
\end{table}

\begin{figure}[ht]
    \centering
    \includegraphics[width=\linewidth]{Figure_G1_alt_qc_designs.png}
    \caption{
    \textbf{Design and optimization results with alternative quasicrystal geometries.}
    (a) Deflation rule for the 4-fold and 8-fold quasicrystal geometries. (b) Quasicrystal-based resonator designs using the same design motif (Fig.~\ref{fig:design_formulation_method}) with 8-fold and 4-fold quasicrystals.(c) Stopband in 8-fold low-mass-contrast design. (d) High-$Q_m$ soft-clamped mode shape in 4-fold and 8-fold designs.
}
    \label{fig:appendix_alt_qc_structures}
\end{figure}

\section{\label{appendix:stopband_validation} Validation of Stopband Simulation with Symmetry-Reduced Model}

\setcounter{figure}{0}
\setcounter{equation}{0}
\setcounter{table}{0}
\renewcommand{\thefigure}{H.\arabic{figure}}
\renewcommand{\theequation}{H.\arabic{equation}}
\renewcommand{\thetable}{H.\arabic{table}}

\begin{figure}[ht]
    \centering
    \includegraphics[width=\linewidth]{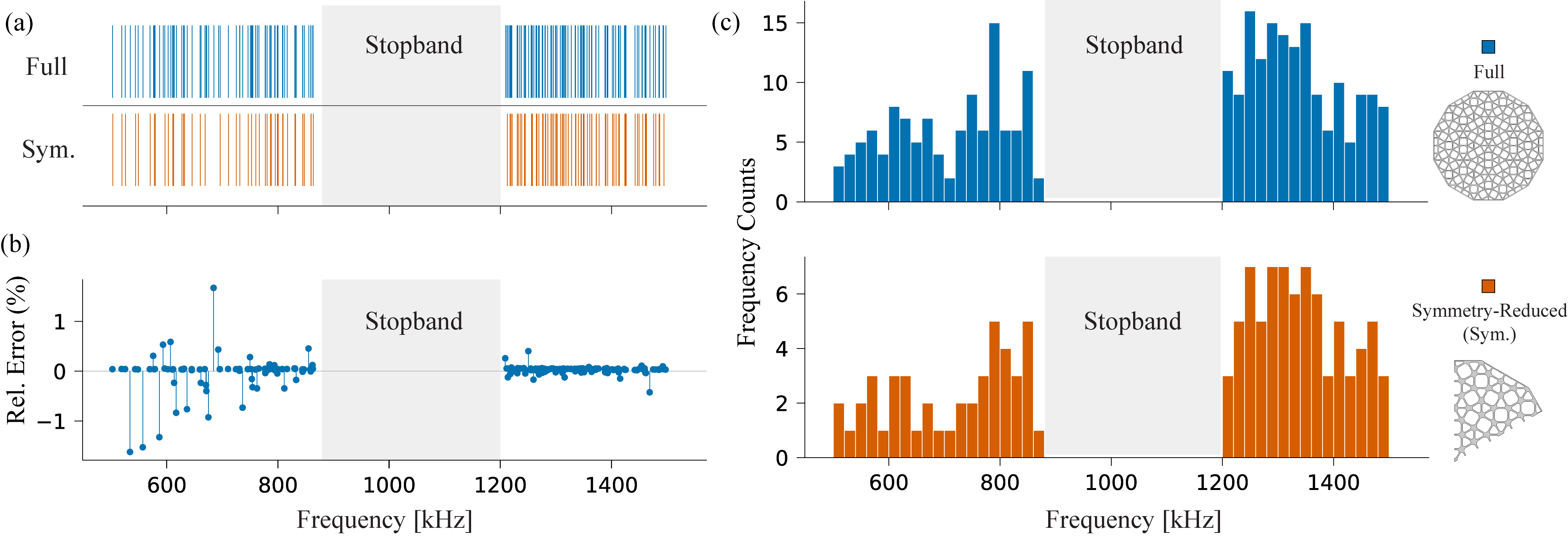}
    \caption{
    \textbf{Eigenfrequency analysis validation: comparison between symmetry-reduced and full-model simulations.}
(a) Mode spectra of the two models. 
(b) Relative frequency differences obtained using nearest-neighbor matching. 
(c) Frequency-count spectra using a 20~kHz bin size.
}
    \label{fig:appendix_stopband_validation}
\end{figure}

As highlighted in Figure~\ref{fig:appendix_simulation_pipeline}, the simulation pipeline relies on a symmetry-reduced model for stopband identification. We validated the stopband identification process in Figure~\ref{fig:bandgap_validation} by demonstrating its performance across different stopband properties. Here, we further validate the eigenfrequency results obtained from the symmetry-reduced simulation by comparing them against a full-model simulation, to assess whether the relevant spectral features are consistently captured by the symmetry-reduced representation. The results are shown in Figure~\ref{fig:appendix_stopband_validation}.

To this end, we performed eigenfrequency analyses of both the symmetry-reduced and full models using identical design parameters, over a frequency range of 500 to 1500~kHz. Figure~\ref{fig:appendix_stopband_validation}(a) presents the resulting mode spectra for both models. To further quantify the discrepancy between the two, we computed the frequency difference between corresponding modes using a nearest-neighbor matching scheme. A strict one-to-one eigenmode correspondence is impractical, owing to the presence of near-degenerate and spurious ("noisy") modes in the full-model results. As shown in Figure~\ref{fig:appendix_stopband_validation}(b), most matched frequencies differ by less than 1\%, with a small number of outliers approaching 2\%, showing close agreement in the locations of the eigenfrequencies obtained from the two models.

We additionally examined the frequency spectrum by tabulating the number of eigenmodes within 20~kHz frequency bins. Although the two models exhibit differences in the absolute mode count per bin, they show similar overall spectral trends. Most importantly, both models identify the same stopband region, indicating that the spectral feature relevant to stopband identification is consistently captured by the symmetry-reduced model. These results further support the use of the symmetry-reduced model for the eigenfrequency analysis underlying stopband identification.

\section{\label{appendix:benchmark_pnc} Benchmarking Performance with Periodic Designs}

\setcounter{figure}{0}
\setcounter{equation}{0}
\setcounter{table}{0}
\renewcommand{\thefigure}{I.\arabic{figure}}
\renewcommand{\theequation}{I.\arabic{equation}}
\renewcommand{\thetable}{I.\arabic{table}}

We compare the proposed QC-based nanomechanical resonator designs with a representative PnC-based design in more detail. For this purpose, we constructed a PnC reference based on the design proposed by Tsaturyan~\textit{et al.}~\cite{tsaturyan_ultracoherent_2017}, using $a=346$~$\mu$m and Mode~E. The QC and PnC designs were simulated in COMSOL using the parameters listed in Table~\ref{tab:appendix_design_param_table}.
The key figures of merit are summarized in Table~\ref{tab:appendix_design_benchmark}. The QC-based resonators show a lower effective mass, particularly for Design 2, leading to the improved thermal-force-noise-limited sensitivity discussed in the main manuscript, although the dissipation dilution factor is lower. The PnC-based design, however, remains a well-established reference with greater design maturity and more extensive theoretical support.

\begin{table}[hb]
\caption{\label{tab:appendix_design_param_table}Design parameters used in the simulation of the QC-based design and the representative PnC-based reference based on~\cite{tsaturyan_ultracoherent_2017}.}
\begin{tabular}{l|c}
Parameters & \\
\hline
Thickness $t$ (nm) & 35 \\
Device dimension (mm) & 3 \\
Young's Modulus (GPa) & 270 \\ 
Density (kg$/m^3$) & 3200 \\
Pre-stress (GPa) & 1.27 \\
\end{tabular}
\end{table}

\begin{table}[hb]
\caption{\label{tab:appendix_design_benchmark}Performance of the QC-based designs (Designs 1 and 2) compared with the PnC-based design.}
\begin{tabular}{l|c|c|c}
\textrm{Designs} & \textrm{1} & \textrm{2} & PnC  \\
\hline
Frequency $f$ (MHz) & 1.29 & 0.88 & 0.78 \\
Stopband Width (MHz) & 0.12 & 0.34 & 0.1 \\
Effective mass $m_\mathrm{eff}$ (ng) & 5.9 & 0.92 & 14.2 \\
Dilution Factor $D_Q$ ($\times 10^3$) & 48.9 & 25.1 & 68.7  \\
\end{tabular}
\end{table}

\section*{Acknowledgments}
The authors gratefully acknowledge Miguel Bessa for helpful discussions and early insights that inspired this work. The authors acknowledge support from the National Research Foundation of Korea (NRF) grant funded by the Korea government (MSIT) (RS-2025-24534529). This work was funded/co-funded by the European Union (ERC, EARS, 101042855). Views and opinions expressed are however those of the author(s) only and do not necessarily reflect those of the European Union or the European Research Council.

\makeatletter
\let\auto@bib@innerbib\@empty
\makeatother

\end{document}